\documentclass[reqno]{article}
\usepackage{ae} % or {zefonts}
\usepackage[T1]{fontenc}
\usepackage[ansinew]{inputenc}
\usepackage{amsmath}
\usepackage{amssymb}
\usepackage{graphicx}
\usepackage[pdftex,usenames,x11names,svgnames,dvipsnames]{xcolor}
\usepackage[colorlinks]{hyperref}
\usepackage{cancel}
\usepackage{enumerate}

\usepackage{tikz}
\usepackage[normalem]{ulem}
\usetikzlibrary{arrows,shapes}
\usepackage{mathdots}
\usepackage{bm} %-------------- Bold Math (AIP RevTex) ------------
\usepackage{mathrsfs} %%------ Special math fonts -----------------

\usepackage{lscape} %landcape pages suppot
\newcommand{\vb}[1]{{\boldsymbol {#1}}}

\newcommand{\ud}{\mathop{}\!\mathrm{d}} %% ----- upright differential d sign------- %%

\newcommand{\ve}{\varepsilon}

\newcommand{\mmr}{M_{\text{reg}}}
\newcommand{\smmr}{m_{\text{reg}}}
\newcommand{\mr}{D}
\newcommand{\TT}{\scriptscriptstyle} %%----------- Smallest size---

\title  {
            Radiation-reaction in classical offshell electrodynamics: 
            \\
            I. The above mass-shell case
        }
\author{    I. Aharonovich  \\
            L. P. Horwitz    }

\begin{document}
\maketitle
\begin{abstract}
    Offshell electrodynamics based on a manifestly covariant off-shell relativistic dynamics of Stueckelberg, 
    Horwitz and Piron, is five-dimensional.
    In this paper, we study the problem of radiation reaction of a particle in motion in this framework.
    
    In particular, the case of above-mass-shell is studied in detail, 
    where the renormalization of the Lorentz force leads to a system of 
    non-linear differential equations for 
    $3$ Lorentz scalars.
        
    The system is then solved numerically, where it is shown that the mass-shell deviation
    scalar $\ve$ either smoothly falls down to $0$
    (this result provides a mechanism for the mass stability of the off-shell theory), 
    or strongly diverges under more extreme conditions.
    In both cases, no runaway motion is observed.
    Stability analysis indicates that the system seems to have chaotic behavior.

    It is also shown that, although a motion under which the mass-shell deviation $\ve$ is constant but not-zero,
    is indeed possible, but, it is unstable,
    and eventually it either decays to $0$ or diverges.
\end{abstract}

\section{Introduction}

Classical 5D electrodynamics arises as a $U(1)$ gauge of the relativistic quantum mechanical Stueckelberg-Schr\"{o}dinger
equation (shown below in \eqref{eq:stueckelberg_schrordinger_evolution}), similar to the construction
of Maxwell fields from the $U(1)$ gauge of the classical Schr\"{o}dinger equation
\cite{Horwitz1998,LandHor1991,LandShnerbHorwitz1995,SaadHorArsh1989}.

The fields depend on the spacetime points $x^{\mu}$ \emph{as well as the invariant evolution parameter $\tau$},
and thus, are defined on a five-dimensional manifold.
In previous papers, we have studied the configuration of these fields associated with a uniformly moving source 
\cite{horjig2006} and uniformly accelerating one \cite{aharonovich_2009}.

In this paper, we continue the investigation of these fields and 
examine the problem of radiation-reaction of a point-particle interacting with its own field, 
using the explicit $\tau$-retarded Green-Function % \eqref{eq:green_function_jigal_2} 
first presented in \cite{aharonovich_2009}.

In this relativistic framework, the mass of the particle $m^2 = - p_{\mu} p^{\mu}$ 
(or its gauge invariant generalization in the presence of electromagnetic fields)
is a \emph{dynamical variable},
whereas $M$, which enters into the Hamiltonian formulation of the theory (see below equation \eqref{eq:stueckelberg_K_function}), 
the so-called \emph{Galilean target mass}\cite{burakovsky_horwitz_1995} (defined as the \emph{on-shell mass}), 
is the non-relativistic limit of $m$.
One then defines a dimensionless \emph{mass-shell deviation scalar} 
\begin{align}
    \label{eq:def_epsilon_mass_definition}
    \ve & = \dfrac{m^2 - M^2}{M^2}
\end{align}
which provides a measure of how far the particle is offshell.

As the fields are essentially 5D, the Huygens' principle no longer applies 
(cf. \cite{GalTsov2002,Kazinski2002,kosyakov2007,mironov-2007}),
and the $\tau$-Green-Function given in ref. \cite{aharonovich_2009} % \eqref{eq:green_function_jigal_2} 
has support on the entire history of the particle, 
within its local 5D past lightcone.
Therefore, the Lorentz force for the self-interacting particle with its own field, 
depends on the particle's own entire history within its past 
lightcone\footnote{In this paper we do not consider external fields.}.

It is shown in section \ref{sec:lorentz_force} that the 
radiation-reaction force differs substantially 
for the three possible cases of $\ve$,
i.e., the \emph{above-mass-shell} case $\ve > 0$, which is the main topic of this paper, 
the \emph{on-mass-shell} case $\ve =0$, and the \emph{below-mass-shell} case $\ve < 0$.
This is mainly due to the $O(4,1)$ choice of symmetry for the homogeneous 
fields\footnote{In \cite{horjig2006}, for example, the choice of initial mass shell for the uniformly moving source 
                was shown to lead to substantially different pre-Maxwell fields, even though their zero-mode 
                corresponds to the same Maxwell fields.}.

For the above-mass-shell case $\ve>0$, it is shown that the force has a renormalizable simple-pole singularity.
After renormalizing this pole, the \emph{residue} is a \emph{local vector equation} with up to $4^{\text{th}}$ order terms in $x^{\mu}$.
After finding an expression for $\ddddot{x}^{\mu}$ which depends only on lower order 
terms $\dot{x}^{\mu}, \ddot{x}^{\mu}$ and $\dddot{x}^{\mu}$,
we are able to construct a set of $6$ coupled scalar equations, with three scalars and their derivatives as dynamical variables,
where $\ve$, the mass-shell-deviation, is the leading scalar.

Apart from the constant $\ve$ case, the analysis to be shown below is numerical (see section \ref{sec:numerical_simulation}), 
performed using the high-precision math library \emph{MPFR} \cite{Fousse:2007:MMP}.
As initial condition, $\ve$ begins with a non-zero positive value, as the particle is above its mass-shell;
it either decreases smoothly towards $\ve \to 0$, or diverges to $\ve \to \infty$.
For the $\ve \to 0$ case, the particle relaxes to an asymptotically on-shell uniform-motion.
In the $\ve \to \infty$ case, the acceleration $\ud^2 x^{\mu} / \ud \tau^2$ increases indefinitely.
It is shown, however, that the \emph{3 velocity} $\ud x^{i} /\ud t$ converges strongly to a finite value, 
such that in any particular Lorentz frame, the particle is asymptotically inertial.

%%
%% add some comments in the results section, regarding how the 3-velocity would seem in different lorentz frames.
%% should be 0 only in a single frame.
%% 

Therefore, in both cases, there is no runaway motion in $\vb{x}(t)$.
Furthermore, the state of motion for $\ve = \text{const} \neq 0$ is shown to be unstable.
%% The non-zero $\ve = \text{const} > 0$, is shown to be unstable. 

For the noninteracting Stueckelberg Hamiltonian, there appears no \emph{a priori} reason for the particle
to have a particular mass. The above-shell analysis of the particle in self-interaction shows that 
for a wide range of conditions converges to the on-shell value. 
This striking result is an important indication of the stability and physical applicability of
the Stueckelberg dynamics.

The remainder of this paper is organized as follows:
%%\begin{enumerate}[I.]
    %%\item 
    In section \ref{sec:fundamentals} an overview is provided of the manifestly covariant relativistic dynamics of Stueckelberg,
          and the corresponding offshell-electrodynamics.
    %%\item 
    In section \ref{sec:lorentz_force} the Lorentz force is formulated using explicit $\tau$-Green-Functions  
    \cite{aharonovich_2009},
    %\eqref{eq:green_function_jigal_2},
    and it is shown how the sign of the mass-shell-deviation $\ve$ leads to substantially different results.
    In this paper, the \emph{above-mass-shell} case is analyzed in detail to provide equations of motion
    (the above mass-shell condition is preserved by the motion).
    We shall deal with the below-mass-shell problem in a later paper.
    %%\item 
    In section \ref{sec:numerical_simulation}, the equations of motion are integrated, and the various results are shown.
    %%\item 
    In section \ref{sec:conclusions}, conclusions of the analysis and simulation are given, with prospects of further investigation.
    %%\item 
    %In the appendices, relations to previously published work on radiation-reaction in this framework is provided.
    Furthermore, the regularization and renormalization procedures of Gel'fand \cite{Gelfand1964_1} are given for reference.
%%\end{enumerate}

% % % % % % % The results reported here are signficantly different from those of previous studies 
% % % % % % % (\cite{OronHorwitz2000} \cite{OronHorwitz2001} \cite{LandHor1991})

\section{Fundamentals}
\label{sec:fundamentals}

\subsection{Relativistic dynamics}
Stueckelberg
postulated a manifestly covariant Hamiltonian-type evolution in both classical 
and quantum mechanics \cite{Stueckelberg1941,Stueckelberg1942}, 
by introducing the Hamiltonian-type evolution 
function for a free 
particle\footnote{Throughout the text, we shall work with standard relativistic notation where $\mu, \nu \in \{ 0,1,2,3\}$,
                  and the Lorentz metric signature has the form $(-1,1,1,1)$, and we take units for which $c=1$},
\begin{align}
    \label{eq:stueckelberg_K_function}
    K & = \dfrac{1}{2M} p_{\mu} p^{\mu}
\end{align}
Phase space is 8D and consists of $x^{\mu} = (t,\vb{x})$ and $p^{\mu} \equiv (E, \vb{p})$, developing with a Lorentz invariant parameter $\tau$.
Particles in this framework trace out worldlines in spacetime, and are therefore, also denoted as \emph{events}.

The corresponding quantum equation of motion is the Stueckelberg-Schr\"{o}dinger equation (we take $\hbar = 1$)
\begin{align}
    \label{eq:stueckelberg_schrordinger_evolution}
    i 
    \dfrac{\partial}{\partial \tau} \psi_{\tau}(x) 
    & =
        K \psi_{\tau}(x)
\end{align}

The framework was extended to a many-body system by Horwitz and Piron \cite{HorPir1973},
where $\tau$ was given a physical significance as a universal historical time, essentially that postulated by Newton.
This interpretation was necessary in order to write equations for the many-body system.
The evolution function $K$ can then take a more general form:
\begin{align}
    \label{eq:stueckelberg_K_function_many_particles_and_potentials}
    K
    & = 
        \sum_{n=1}^{N}
            \dfrac{1}{2M_{n}}
            \eta_{\mu \nu}
            p^{\mu}_{n} 
            p^{\nu}_{n}
        +
        V(x_{1}, x_{2}, \ldots, x_{N})
\end{align}
$V(x_1, x_2, \ldots ) \equiv V(x)$ is a Lorentz scalar function of spacetime positions of all the particles at the (universal) time $\tau$,
which establishes a dynamical correlation between them.
The classical equations of motion are similar to those of the non-relativistic Hamilton equations
\begin{align}
    \label{eq:hamilton_equations_of_stueckelberg}
    \dot{x}^{\mu}_{n}
    & = 
        \dfrac{\partial K}{\partial p_{n \, \mu}}
    =
        \dfrac{1}{M_{n}} p^{\mu}_{n}
    &
    \dot{p}_{n \, \mu}
    & = 
        - \dfrac{\partial K}{\partial x^{\mu}_{n}}
    =
        - \dfrac{\partial V}{\partial x^{\mu}_{n}}
\end{align}
In the usual application of special-relativity (SR, cf. \cite{Rindler1991,weinberg_1972}), the energy-momentum is constrained to a
\emph{mass-shell} defined as:
\begin{align}
    \label{eq:ordinary_sr_mass_shell_definition}
    p^{\mu} p_{\mu} & = \vb{p}^2 - E^2 = -M^2
\end{align}
where $M$ is a given \emph{fixed quantity}, a property of the particle.
In the Stueckelberg formulation, however, the particle's mass is a dynamical property which may depend on $\tau$:
\begin{align}
    \label{eq:mass_shell_definition}
    p^{\mu} p_{\mu} & = - m^2
\end{align}
The relation between $\tau$ and \emph{proper time} $s$ is given, according to \eqref{eq:hamilton_equations_of_stueckelberg}, by
\begin{align}
    \label{eq:proper_time_vs_tau}
    \ud s_{n}^2 
    & \equiv
        - \ud x_{n}^{\mu} \ud {x}_{n \, \mu}
    =
        - 
        \dot{x}_{n}^{\mu} \dot{x}_{n \, \mu}
        \ud \tau^2
    =
        - 
        \dfrac{1}{M^2} p_{n}^{\mu} p_{n \, \mu}
        \ud \tau^2
    =
        \dfrac{m_{n}^2}{M_{n}^2} \ud \tau^2
\end{align}
Thus, the proper time interval $\ud s_{n}$ and the universal time interval $\ud \tau$, are related
through the ratio between the dynamical Lorentz invariant mass $m_n$, and the so called 
\emph{Galilean target mass} $M_n$.
If $V(x) \to 0$ asymptotically for large $\tau$, then each $m_n^2$ becomes a constant,
since $\ud p_{n}^{\mu}/\ud \tau \to 0$, we find
\begin{align*}
    %\lim\limits_{\tau \to \infty}
    K 
    & = 
        \sum_{n}
        \dfrac{1}{2M_{n}} 
        p_{n}^{\mu} p_{n \, \mu}
    =
        - 
        \sum_{n}
        \dfrac{m_{n}^2}{2M_{n}}
    =
        \text{const}
\end{align*}
Since this asymptotic value is what we expect to measure in an experiment, we may take
$M_n$ to be this asymptotic value.
Is has been an open question in the framework of this theory
as to why a given particle appears (even approximately) 
to have the same value of $m$ in many circumstances,
e.g., after many scatterings.
Possible answers lie in the properties of a particle in interaction with other 
particles\footnote{T. Jordan, personal communication.}, 
or a relaxation to a minimal free-energy.
Some insight has been gained in the framework of relativistic mechanics \cite{HorwitzSchievePiron1981}
in the discovery of a high-temperature phase transition for an offshell particle to become confined to a particular 
mass shell \cite{burakovsky_horwitz_schieve_1996}.
In this work, a new, \emph{renormalized mass}  $\mmr$ emerges, related to $M$ by some very large scale, 
and a new, very natural and intrinsic mechanism is found, in which the self-interaction
of a relativistic charged particle can cause a relaxation of the particle mass to its \emph{renormalized mass}.
A possible relation between these results will be studied elsewhere.

% % % A mechanism has been found  \cite{burakovsky_horwitz_schieve_1996}  in a possible high-temperature
% % % phase transition for an offshell particle to become restricted to a particular value.
% % % In this work, we show that the self-interaction of a relativistic charged particle can cause
% % % a relaxation of the particle mass to a mass-shell defined by $M$.

% % % In this paper, it is shown that radiation-reaction presents 
% % % a dynamical mechanism in which 
% % % a particle falls back to its mass-shell limit.

% % A different, statistical argument was used \cite{burakovsky_horwitz_schieve_1996} to obtain
% % a similar result, however \emph{only as a non-relativistic limit}.

% ================================================================================================================= %
% ----------------------------------------------------------------------------------------------------------------- %
% ----------------------> OSE Electrodynamics <-------------------------------------------------------------------- %
% ----------------------------------------------------------------------------------------------------------------- %
% ================================================================================================================= %
\subsection{Off-Shell Electrodynamics}
What we shall call "pre-Maxwell" off-shell electrodynamics is
constructed in a similar fashion to the formal construction of standard
Maxwell electrodynamics from the Schr\"{o}dinger equation
\cite{SaadHorArsh1989} (see also \cite{PhysRevLett.33.445,NYAS:NYAS86}).

Under the local gauge transformation
\begin{align}
    \label{eq:gauge_transformation_definition}
    \Psi'_{\tau}(x) = e^{- i e_0 \chi(x,\tau)} \Psi_{\tau}(x)
\end{align}
5 compensation fields $a^{\alpha}(x,\tau)$ ($\alpha \in
\{0,1,2,3,5\}$) are implied by \eqref{eq:stueckelberg_K_function} and \eqref{eq:stueckelberg_schrordinger_evolution}, 
such that with the transformation
\begin{align*}
    a'_{\TT \alpha}(x,\tau) = a_{\alpha}(x,\tau) - \partial_{\alpha} \chi(x,\tau)
\end{align*}
in the following modified Stueckelberg-Schr\"odinger equation remains form invariant
\begin{align}
    \label{eq:stueckelberg_schrodinger_after_gauge}
    \left[ i \dfrac{\partial}{\partial \tau} + e_0 a_{5}(x,\tau) \right] \Psi_{\tau}(x) =
    \dfrac{1}{2M}
    \left[
        (p^{\mu} - e_0 a^{\mu})
        (p_{\mu} - e_0 a_{\mu})
    \right]
    \Psi_{\tau}(x)
\end{align}
under the transformation \eqref{eq:gauge_transformation_definition}.

The resulting $K$ evolution function is then of the same form 
as for the usual $U(1)$ gauge
compensation argument for the non-relativistic Schr\"{o}dinger
equation (where the fourth field $a^{0}$ arises from the requirement for the gauge invariance for the time 
dependent Schr\"{o}dinger-equation), here given by:
\begin{align}
    K =
        \dfrac{1}{2M}
        \left[
            p - e_0 a(x,\tau)
        \right]^2
        -
        e_0 a^{5}(x,\tau)
\end{align}
(where we have used the shorthand notation of $x^2 = x_{\mu} x^{\mu}$),
and the corresponding Hamilton equations are
\begin{align}
    \label{eq:Hamilton_Equations_1}
    \dot{x}^{\mu}(\tau) & =
        \dfrac{\partial K}{\partial p_{\mu}} =
        \dfrac{1}{M}
        \left[
            p^{\mu} - e_0 a^{\mu}
        \right] \\
    \label{eq:Hamilton_Equations_2}
    \dot{p}^{\mu}(\tau) & =
        - \dfrac{\partial K}{\partial x_{\mu}} =
        \dfrac{e_0}{M}
        \left(
            p - e_0 a(x,\tau)
        \right)_{\nu}
        \partial^{\mu} a^{\nu}(x,\tau) +
        e_0 \partial^{\mu} a^{5}(x,\tau)
\end{align}
Here, $e_0$ is proportional to the Maxwell charge $e$ through a
dimensional constant, which is discussed below. Second order
equations of motion for $x^{\mu}(\tau)$, a generalization of the
usual Lorentz force, follow from the Hamilton equations
\eqref{eq:Hamilton_Equations_1} and \eqref{eq:Hamilton_Equations_2}
\begin{align}
    \label{eq:5D_Lorentz_force}
    M \ddot{x}^{\mu} = e_0 \dot{x}^{\nu} {f^{\mu}}_{\nu} + e_0 {f^{\mu}}_{5}
\end{align}
where for $\alpha,\beta = 0,1,2,3,5$ the antisymmetric tensor
\begin{align}
    \label{eq:ose_fields_from_potentials}
    f^{\alpha \beta} \equiv \partial^{\alpha} a^{\beta} - \partial^{\beta} a^{\alpha}
\end{align}
is the (gauge invariant) 5D field tensor. Moreover, second order
wave equation for the fields $f^{\alpha \beta}$ can be derived from
a Lagrangian density 
\begin{align}
    \label{eq:ose_field_langarangian}
    \mathscr{L} =  -\dfrac{\lambda}{4} f_{\alpha \beta} f^{\alpha \beta} - e_0 a_{\alpha} j^{\alpha}
\end{align}
which produces, by the usual variational methods, the wave equation
\begin{align}
    \label{eq:ose_fields_wave_equation}
    \lambda \partial_{\alpha} f^{\beta \alpha} = e_0 j^{\beta}
\end{align}
$\lambda$ is a dimensional constant, which will be shown below to have dimensions of length.
The sources $j^{\beta}(x,\tau)$ depend both on spacetime and on $\tau$, and obey
the continuity equation
\begin{align}
    \label{eq:continuity}
    \partial_{\alpha} j^{\alpha} = \partial_{\mu} j^{\mu} + \partial_{\tau} \rho = 0
\end{align}
where $j^{5} \equiv \rho$ is a Lorentz invariant \emph{spacetime
density of events}. This equation follows from
\eqref{eq:stueckelberg_schrodinger_after_gauge} for
\begin{align*}
    \rho_{\tau}(x) & = \Psi^{*}_{\tau}(x) \Psi_{\tau}(x)
    \\
    j^{\mu}_{\tau}(x) & =
        - \dfrac{i}{2M}
        \left[
            \Psi^{*}_{\tau}(x)
            \left(
                i \partial^{\mu} - e_{0} a^{\mu}(x,\tau)
            \right)
            \Psi_{\tau}(x)
            +
            c.c.
        \right]
\end{align*}
as we discuss below, and also from the classical argument given
below.

\subsubsection{Currents of point events}
Jackson \cite{Jackson1995} showed that a conserved current for a
moving point charge can be derived in a covariant way by defining
the current as
\begin{align}
    \label{eq:maxwell_current_of_point_particle}
    J^{\mu}(x) = e \int_{-\infty}^{+\infty} \ud s \, \dot{z}^{\mu}(s) \delta^4 [x - z(s)]
\end{align}
In this case, $s$ is the proper time, and $z^{\mu}(s)$ the
world-line of the point charge (for free motion, $s$ may coincide
with $\tau$), and $\dot{z}^{\mu}(s) = \dfrac{d}{ds} z^{\mu}(s)$.
Then,
\begin{align}
    \label{eq:maxwell_current_conservation_proof}
    \partial_{\mu} J^{\mu} =
        - e \int_{-\infty}^{+\infty} \ud s \, \dfrac{\ud}{\ud s} \delta^4 [x - z(s)] =
        - e \, \lim \limits_{L \rightarrow +\infty} \delta^4 [x - z(s)] \Bigg|_{-L}^{+L}
\end{align}
which vanishes if $z^{\mu}(s)$ (or, for example, just the time
component $z^{0}(s)$) becomes infinite for $s \rightarrow \pm
\infty$, and the observation point $x^{\mu}$ is restricted to a
bounded region of spacetime, e.g., the laboratory. We therefore,
with Jackson (see also Stueckelberg \cite{Stueckelberg1941}), identify $J^{\mu}$ as the Maxwell current. We see that
this current is a \emph{functional} on the world line, and the usual
notion 
of a "particle" associated with a conserved 4-current (and 
therefore a conserved charge corresponding to the space integral of the 
fourth component), 
corresponds to this functional on the world line.

If we identify $\delta^4 [x - z(s)]$ with a density $\rho(x,s)$
and the local (in $\tau$) current 
$\dot{z}^{\mu}(s) \delta^4 [x -z(s)]$ with a local current $j^{\mu}(x,s)$
\begin{align}
    \label{eq:ose_current_of_point_event}
    \rho(x,s)    & = \delta^4 [x - z(s)] \qquad \qquad
    &
    j^{\mu}(x,s) & = \dot{z}^{\mu}(s) \delta^4 [x - z(s)]
\end{align}
then the relation
\begin{align*}
    \dfrac{d}{ds} \delta^4 [x - z(s)] = -\dot{z}^{\mu}(s) \partial_{\mu} \delta^4 [x - z(s)]
\end{align*}
used in the above demonstration in fact corresponds to the
conservation law (reverting to the more general parameter $\tau$ in
place of the proper time $s$) \eqref{eq:continuity}
\begin{align}
    \label{eq:ose_current_conservation}
    \partial_{\mu} j^{\mu}(x,\tau) + \partial_{\tau} \rho(x,\tau) = 0
\end{align}
What we call the \emph{pre-Maxwell} current of a point \emph{event}
is then defined as
\begin{align}
    \label{eq:point_source_current}
    j^{\alpha}(x,\tau) = \dot{z}^{\alpha}(\tau) \delta^4 [x - z(\tau)]
\end{align}
where $j^{5}(x,\tau) \equiv \rho(x,\tau)$ and $\dot{z}^{5}(\tau)
\equiv 1$ (since $z^{5}(\tau) \equiv \tau$). 
Note that $j^{\alpha}$ \emph{does not} have $O(4,1)$ tensor properties, 
since the factor $\dot{x}^{\mu}$ in $j^{\mu}$ cannot be obtained from $j^{5}$ by a linear transformation
(this also occurs in the 4D theory where the current is constructed from a $\delta^3(x)$ for the $J^{0}$ component
and the three-vector components $J^{i}$ with velocity times this 
$\delta^3(x)$\footnote
             {
                 The covariant Maxwell theory assumes $J^{\mu}(x)$ to be a given $4$-vector
                 (not based on the Jackson type construction \cite{Jackson1995})
                 and therefore, the theory is of course, Lorentz covariant. We have no \emph{a priori}
                 reason however, to assume $O(4,1)$ covariance of the theory.
             }
),
and therefore, the fields associated with this current are not tensors under $O(4,1)$;
the components $a^{\mu}$ are, however, tensors under $O(3,1)$, the Lorentz group.

Integrating
\eqref{eq:ose_fields_wave_equation} over $\tau$, we recover the
standard Maxwell equations for Maxwell fields defined by
\begin{align}
    A^{\mu}(x) & = \int a^{\mu}(x,\tau) \ud \tau
    \label{eq:A_from_a}
\end{align}
We therefore call the fields $a^{\mu}(x,\tau)$ \emph{pre-Maxwell
fields}. We see this by noting \eqref{eq:ose_field_langarangian}
\begin{align}
    \lambda
    \int_{-\infty}^{+\infty}
        \left[
            \partial_{\mu} f^{\mu \nu}(x,\tau)
            +
            \partial_{5} f^{\mu 5}(x,\tau)
        \right]
        \ud \tau
    & =
        \lambda
        \int_{-\infty}^{+\infty}
            \partial_{\alpha} f^{\mu \alpha}(x,\tau)
        \ud \tau
    \nonumber
    \\
    & =
        e_{0} \int_{-\infty}^{+\infty} j_{\beta}(x,\tau) \, \ud \tau
    \nonumber
    \\
    & =
        e_{0} \dfrac{1}{e} J^{\mu}(x)
    \label{eq:derivation_of_A_from_a}
\end{align}
where the $\tau$ integral of $\partial_{5} f^{\mu 5}$ is assumed to vanish.
This can be accounted for by considering the field $f^{\mu 5}$ 
as a wave-packet that propagates to infinity in space time for $\tau \rightarrow \infty$,
or,
in terms of its Fourier transform to its offshell mass distribution,
\begin{align}
    \label{eq:a_mu_fourier_transform}
    \tilde{a}^{\mu}(x,\kappa) = \int_{-\infty}^{+\infty} e^{i\kappa\tau} a^{\mu}(x,\tau) \, \ud \tau,
\end{align}
as an application of the Riemann-Lebesgue lemma. Eq. \eqref{eq:A_from_a} then implies that the Maxwell
potentials and fields correspond to the \emph{zero-mode} of the pre-Maxwell offshell fields,
with respect to $\tau$, i.e.,
\begin{align}
    \label{eq:maxwell_field_from_zero_mode}
    A^{\mu}(x) = \tilde{a}^{\mu}(x,\kappa) |_{\kappa = 0}
\end{align}

% % % or understood in terms of asymptotic $\tau$ independence of the fields (approach to the zero mode).  
% % % We shall discuss this point further below. 

The relation \eqref{eq:A_from_a} then follows. Moreover, since
\begin{align}
    \partial_{\nu} F^{\mu \nu}
    & =
        \dfrac{e_{0}}{\lambda e}
        J^{\mu}(x),
\end{align}
the constants $\lambda$ and $e_0$ are related by
\begin{align}
    \lambda & = \dfrac{e_{0}}{e}
    &
    e       & = \dfrac{e_{0}}{\lambda}
\end{align}
where $e$ is the standard Maxwell charge.

For the quantum theory, a real positive definite density function $\rho_{\tau}(x)$
can be derived from the Stueckelberg-Schr\"odinger equation \eqref{eq:stueckelberg_schrordinger_evolution}
\begin{align}
    \rho(x,\tau) = |\Psi_{\tau}|^2 = \Psi^{*}_{\tau} (x) \Psi_{\tau}(x)
\end{align}
which goes over to $\rho(x,\tau) = \delta^4[x -
z(\tau)]$ in the classical (non-quantum) limit. It follows from the
Stueckelberg equation
\eqref{eq:stueckelberg_schrodinger_after_gauge} that the continuity
equation \eqref{eq:ose_current_conservation} is then satisfied for
the gauge invariant current
\begin{align}
    j^{\mu}(x,\tau) =
        - \dfrac{1}{2M}
        \left[
            \Psi^{\ast}_{\tau}(x) (i \partial^{\mu} - e_0 a^{\mu}(x,\tau)) \Psi_{\tau}(x)  +
            \text{c.c.}
        \right]
\end{align}
Using \eqref{eq:ose_field_langarangian}, applicable in the quantum
theory as well \cite{LandShnerbHorwitz1995,SaadHorArsh1989}, and
\eqref{eq:A_from_a}, we find
\begin{align}
    \label{eq:ose_maxwell_current_integral}
    J^{\mu}(x) = e \int_{-\infty}^{+\infty} j^{\mu}(x,\tau) \, \ud \tau
\end{align}

\subsubsection{The wave equation}
From equations \eqref{eq:ose_fields_wave_equation} and \eqref{eq:ose_fields_from_potentials} one can derive
the wave equation for the potentials $a^{\TT\alpha}(x,\tau)$:
\begin{align}
    \label{eq:ose_potentials_wave_equation_before_gauge}
    \lambda \partial_{\beta} \partial^{\beta} a^{\alpha} -
    \lambda \partial^{\alpha} (\partial_{\beta} a^{\beta}) = e_0 \, j^{\alpha}
\end{align}
Under the generalized Lorentz gauge $\partial_{\beta} a^{\TT\beta} =
0$, the wave equation takes the simpler form
\begin{align}
    \label{eq:ose_potentials_wave_equation_after_gauge}
    \lambda \partial_{\beta} \partial^{\beta} a^{\alpha} =
    \lambda \left[ \Box^2 a^{\alpha} + \sigma_5 \dfrac{\partial^2 a^{\alpha}}{\partial \tau^2} \right] =
    e_0 \, j^{\alpha}(x,\tau)
\end{align}
where the $5^{th}$ diagonal metric component can take either sign 
$\sigma_5 = \pm 1$, corresponding to $O(4,1)$ and $O(3,2)$
symmetries of the homogeneous field equations, respectively.

Integrating \eqref{eq:ose_potentials_wave_equation_after_gauge} with respect to $\tau$,
and assuming, as above, that $\lim\limits_{\tau \rightarrow \pm \infty} a^{\alpha}(x,\tau) =0$
we obtain
\begin{align*}
    \lambda
    \int_{-\infty}^{+\infty} \ud \tau \,
    \left[ \Box^2 a^{\alpha} + \sigma_5 \dfrac{\partial^2 a^{\alpha}}{\partial \tau^2} \right] =
    \dfrac{e_0}{e} J^{\alpha}(x)
\end{align*}

% Assuming that $\lim\limits_{\tau \to \pm \infty} \partial_{\tau} a^{\alpha}(x,\tau) = 0$, 
With \eqref{eq:A_from_a}, we obtain
\begin{align}
    \label{eq:4D_wave_equation_after_tau_integration}
    \lambda \Box^2 A^{\mu}(x) & = \dfrac{e_0}{e} J^{\mu}(x)
    \\
    \intertext{i.e.}
    \Box^2 A^{\mu}(x) & = J^{\mu}(x)
\end{align}
(for $\mu = 0,1,2,3$)

Therefore, the Maxwell electrodynamics is properly contained in the
5D electromagnetism.

\subsubsection{A note about units}
In natural units ($\hbar = c = 1$), the Maxwell potentials $A^{\mu}$ have units of $1/L$.
Therefore, the pre-Maxwell offshell potentials $a^{\alpha}$ have units of $1/L^2$, and
in order to maintain the action integral
\begin{align}
    S = \int_{-\infty}^{+\infty} \mathscr{L} \, \ud \tau \, \ud^4x
\end{align}
dimensionless, the coefficient $\lambda$ in \eqref{eq:ose_field_langarangian} must have units of $L$, forcing
$e_0$ to have units of $L$ as well ($e$ is dimensionless).

% ================================================================================================================= %
% ----------------------------------------------------------------------------------------------------------------- %
% ----------------------> Solutions of the equations <------------------------------------------------------------- %
% ----------------------------------------------------------------------------------------------------------------- %
% ================================================================================================================= %
\subsection{Solutions of the wave equation}
In our previous works \cite{horjig2006,aharonovich_2009}, we have discussed the Green-functions
associated with the 5D wave-equation \eqref{eq:ose_potentials_wave_equation_after_gauge}.
In particular, using the ultrahyperbolic Riemann-Liouville integro-differential operator of Nozaki \cite{Nozaki_1964},
the $\tau$-retarded form was shown in to be (for the $O(4,1)$ case, which we shall concentrate on here, $\sigma_{5} = +1$):
\begin{align}
    \label{eq:tau_retarded_green_function}
    g(x,\tau)
    & = 
        -
        \dfrac{\theta(\tau)}{4\pi^2}
        \dfrac{\theta( - x^2 - \tau^2 ) }
              {[ - x^2 - \tau^2 ]^{3/2} }
\end{align}
where:
\begin{align*}
    - x^2 - \tau^2
    & = 
        - x_{\mu} x^{\mu} - \tau^2 
    \equiv
        - x_{\alpha} x^{\alpha}
\end{align*}

Thus, the fields of a point source can be directly integrated to yield:
\begin{align}
    \label{eq:hps_potential_solution_using_green_function}
    a^{\alpha}(x,\tau) =
            e_0
            \int \ud^4x' \, \ud \tau'
                \,
                g(x-x',\tau-\tau') \, j^{\alpha}(x',\tau')
\end{align}
and applying it to a point particle represented by \eqref{eq:ose_current_of_point_event}.
The potentials $a^{\alpha}(x,\tau)$ are then given by:
\begin{align}
    \label{eq:hps_potential_solution_using_green_function_for_point_particle}
    a^{\alpha}(x,\tau) 
    & =
            e_0
            \int_{-\infty}^{+\infty}
                \ud \tau'
                \,
                g(x-z(\tau'),\tau-\tau') \, \dot{z}^{\alpha}(\tau')
    \nonumber
    \\
    & = 
            -
            \dfrac{e_0}{4\pi^2}
            \mathcal{R}
            \int_{-\infty}^{\tau}
                \dfrac
                %% numerator %%
                {
                    \theta [ - (x - z(\tau'))^2 - (\tau - \tau')^2  ]
                }
                %% denominator %%
                {
                    [ - (x - z(\tau'))^2 - (\tau - \tau')^2 ]^{3/2}
                }
                \,
                \dot{z}^{\alpha}(\tau')
                \,
                \ud \tau'
\end{align}
where the $\mathcal{R}(\ldots)$ indicates that the integration
\eqref{eq:hps_potential_solution_using_green_function_for_point_particle} 
is \emph{regularized} at the points where the denominator vanishes,
which, by the $\theta(\ldots)$ function in the numerator, are also \emph{the bounds of the integration}.
We shall explain this regularization procedure in detail in the following.

The $\tau$-retarded Green-Function \eqref{eq:tau_retarded_green_function} 
leads to $\tau$-retarded solutions of the wave-equation.
However, their \emph{zero-mode}, identified with the Maxwell field \eqref{eq:maxwell_field_from_zero_mode},
is not $t$-retarded, since the $\tau$-integration of 
the fields \eqref{eq:hps_potential_solution_using_green_function_for_point_particle}
are, essentially, symmetric under time $t$-reversal, and results in the average of $t$-retarded and $t$-advanced potentials.
This average does not contribute to the Maxwell radiation field \cite{dirac_1938},
and is singular in the zero-mode.
It is, however, propagating on the pre-Maxwell level and well defined by \eqref{eq:hps_potential_solution_using_green_function_for_point_particle}.

Equation \eqref{eq:hps_potential_solution_using_green_function_for_point_particle} above is the foundation
on which the Lorentz force of back-reacting particle is evaluated in the next section.

\section{Lorentz force}
\label{sec:lorentz_force}
The potentials $a^{\alpha}(x,\tau)$ \eqref{eq:hps_potential_solution_using_green_function_for_point_particle}
can be used to find the corresponding fields ${f^{\alpha}}_{\beta}$,
and with them, the Lorentz force \eqref{eq:5D_Lorentz_force}.

We then find the radiation-reaction equation to be 
\begin{align}
    \label{eq:radiation_reaction_1}
    M \ddot{x}^{\mu}(\tau)
    & =
        (-1)^2
        \dfrac{e^2_0}{4\pi^2}
        \dfrac{3}{2}
        \dot{x}_{\alpha}(\tau)
        \mathcal{R}
        \int_{-\infty}^{\tau}
            \theta ( R(\tau,\tau') ) 
            \dfrac{ h^{\alpha \mu}(\tau,\tau')  }
                  { R^{5/2}(\tau,\tau')         }
            \ud \tau'
\end{align}
where
\begin{align}
    \label{eq:def_R}
    R(\tau,\tau')
    & = 
        - (x_{\alpha}(\tau) - x_{\alpha}(\tau')
          (x^{\alpha}(\tau) - x^{\alpha}(\tau')
    =
    \nonumber
    \\
    & = 
        - 
        (x_{\mu}(\tau) - x_{\mu}(\tau'))
        (x^{\mu}(\tau) - x^{\mu}(\tau'))
        - 
        (\tau - \tau')^2
\end{align}

The integration is bounded by regions where $R(\tau, \tau')$ is positive within the range $-\infty<\tau' < \tau$,
i.e., \emph{within the past 5D light cone} of the particle at $x^{\alpha}(\tau)$.

The numerator function $h^{\alpha \beta}(\tau,\tau')$ is given by:
\begin{align}
    \label{eq:def_h_alpha_mu}
    h^{\alpha \beta}(\tau,\tau')
    & = 
        \dot{x}^{\alpha}(\tau') \dfrac{\partial}{\partial x_{\beta }(\tau)} R(\tau,\tau')
        -
        \dot{x}^{\beta }(\tau') \dfrac{\partial}{\partial x_{\alpha}(\tau)} R(\tau,\tau')
\end{align}

Defining $h = \tau - \tau'$, we expand $R(\tau,\tau') = R(\tau, \tau-h)$ around $h=0$:
\begin{align}
    \label{eq:R_h_expansion}
    R(\tau,\tau')
    & = 
         - 
         h^2 
         \dot{x}_{\alpha} \dot{x}^{\alpha}
         + 
         h^3 \dot{x}_{\alpha} \ddot{x}^{\alpha}
         - 
         \dfrac{h^4}{4}
         \ddot{x}_{\alpha} \ddot{x}^{\alpha}
         - 
         2 
         \dfrac{h^4}{3!}
         \dot{x}_{\alpha} \dddot{x}^{\alpha}
         +
         O(h^{5})
\end{align}
Similarly, expanding $h^{\alpha\beta}(\tau,\tau-h)$ around $h=0$:
\begin{align}
    \label{eq:h_h_expansion}
    h^{\alpha \beta}(\tau,\tau')
    & = 
        h^2 ( \dot{x}^{\alpha} \ddot{x}^{\beta} - \dot{x}^{\beta} \ddot{x}^{\alpha})
        + 
        O(h^3)
\end{align}

Let us define the \emph{leading term in $R(\tau,\tau')$} as $\ve$:
\begin{align}
    \label{eq:mass_shell_definition_x_dot_x_dot}
    \ve(\tau) 
    & = 
        - \dot{x}_{\alpha} \dot{x}^{\alpha}
    =
        - 
        \dot{x}_{\mu} \dot{x}^{\mu}
        - 
        1
\end{align}
where we have used the definition of $x^{5}(\tau) \equiv \tau$ and $\dot{x}^{5}(\tau) \equiv 1$.
To see how this coincides with the definition given in the introduction \eqref{eq:def_epsilon_mass_definition},
let us write:
\begin{align*}
    \dot{x}_{\mu}  \dot{x}^{\mu}
    & = 
        -
        \dfrac{(p_{\mu} - e_0 a_{\mu})(p^{\mu} - e_0 a^{\mu})}{M^2}
    =
        - 
        \dfrac{m^2}{M^2}
\end{align*}
and thus, clearly:
\begin{align*}
    \ve
    & = 
         - 
         \dot{x}_{\mu} \dot{x}^{\mu}
         -
         1
    =
        \dfrac{m^2}{M^2}
        -
        1
    =
        \dfrac{m^2 - M^2}{M^2}
\end{align*}

\begin{figure}
    \begin{center}
        \includegraphics[width=12cm]{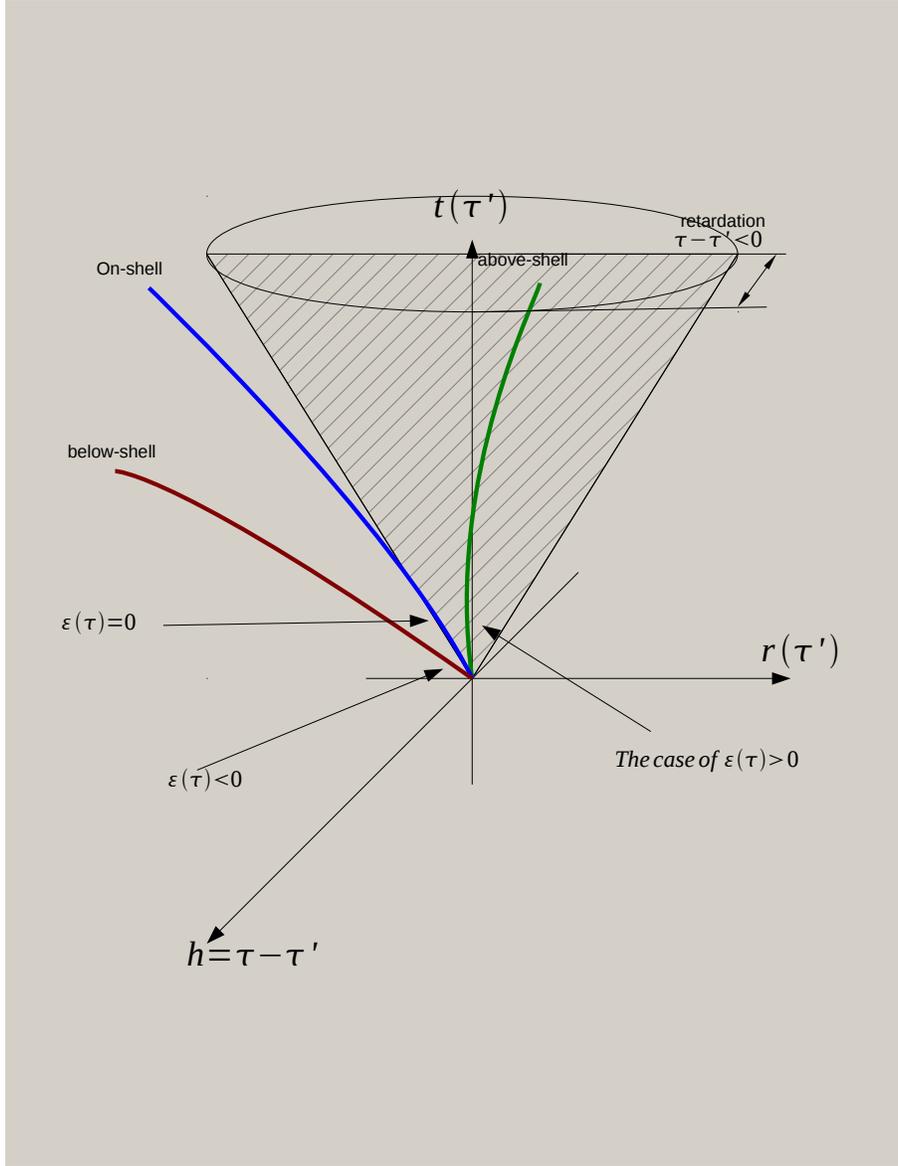}
    \end{center}
    \caption{
                The past light cone, where the vertex in the origin is the present point $x^{\alpha}(\tau)$ 
                (only $t, r$ and $\tau$ axes are plotted). The 5D light cone interior is given by 
                   $(t(\tau) - t(\tau'))^2 \geq (\tau - \tau')^2 + (\vb{r}(\tau') - \vb{r}(\tau))^2$,
                   where the axis for $h = \tau - \tau'$ is drawn.
                   Retardation is when $h > 0$, bounded by the hatched surface at $h=0$.
                   $3$ possible trajectories are shown: inside the past lightcone, plotted in \textcolor{green!30!black}{green},
                   corresponding to the $\ve(\tau) > 0$ case, where $\ve(\tau)$ is essentially the slope $\dot{t}(\tau)^2 - \dot{r}^2 - 1$.
                   The case of $\ve(\tau) < 0$ is shown in \textcolor{red!60!black}{red} and $\ve(\tau) = 0$ in \textcolor{blue}{blue}.
            }
    \label{fig:light_cone_and_paths}
\end{figure}

The function $R(\tau,\tau')$ is the \emph{generalized Lorentzian distance function} (interval on the $4+1$ space) 
between $2$ points in the particle's history, namely, its \emph{present location} at $x^{\alpha}(\tau)$
and its \emph{previous} location at $x^{\alpha}(\tau')$ at the time $\tau'$.
The distance converges to $R(\tau,\tau') \to 0$ 
as $\tau' \to \tau$, or as $h = \tau - \tau' \to 0^{+}$.
However, the \emph{sign} of $\ve$ at $\tau$ determines whether the \emph{immediate past} is 
inside the domain of integration:
\begin{itemize}
    \item We call the case of $\ve(\tau) > 0$ the \emph{above mass-shell case}. In this case,
          the integration runs up to the point $\tau = \tau'$ itself, where 
          there is a strong singularity. 
          From the leading terms in $R(\tau,\tau-h)$ and $h^{\alpha \beta}(\tau,\tau-h)$, 
          we see that the singularity is of the order of $2 \cdot 5/2 - 2 = 3$.
          
    \item $\ve(\tau) < 0$ is the \emph{below mass-shell case}, where the integration is bounded by some $\tau_1 < \tau$,
          where $R(\tau,\tau_1) = 0$ and $R(\tau, \tau - h) < 0$ for all $0 < h < \tau - \tau_1$.
          I.e., in such a case, the particle is \emph{5D spacelike} with respect to its immediate past, up to $\tau_1$,
          and therefore, that part of the history does not influence, or back react, with the particle at its present position.
          
          The interaction is only with parts of the particle's history which are \emph{inside the 5D past cone}
          with respect to its present 5D position $x^{\alpha}(\tau)$.
          
          In the neighborhood of these \emph{crossing points}, denoted by $\tau_i$, the expansion $R(\tau,\tau_i)$ takes the form:
          \begin{align}
              %\label{eq:R_expansion_at_past_points}
              R(\tau, \tau_i - h)
              & =
                  h 
                  \dot{R}(\tau_i)
                  + 
                  O(h^2)
          \end{align}
          I.e., for the general case, $\dot{R}(\tau_i) \neq 0$, i.e., where the motion in and out of the past light-cone is well defined, 
          the singularity due to the vanishing denominator is of order $5/2$, namely, non-integral order, 
          which enables regularization to remove it.
          
    \item $\ve(\tau) = 0$, is the \emph{on mass-shell case},  $m^2 = M^2$.
          In this case, higher order terms determine whether the orbit \emph{in the immediate past} $\tau' \to \tau$ 
          is inside the past 5D cone or not.
          Locally, at $\tau' = \tau$, the particle's orbit is \emph{tangential to the 5D cone}.
          If this is persistent, i.e., if the on-mass-shell motion existed over some measurable interval 
          in the particle's history, and the immediate past is \emph{inside the lightcone}, 
          then the singularity of the denominator is even stronger, at least of order $4 \cdot 5/2 - 2 = 8$.
\end{itemize}
The various possible trajectories are shown in figure \ref{fig:light_cone_and_paths}, where, in this case, the on-shell
motion is actually outside the lightcone, then becoming tangential at $\tau' = \tau$, or $h = 0$, at the vertex of the cone at $x^{\mu}(\tau)$.

Henceforth, we shall study the \emph{above mass-shell} case, and thus, assume $\ve(\tau) > 0$ for 
all the range of relevant $\tau$.
The other cases will be studied in a succeeding publication.

Let us begin by breaking up the integration \eqref{eq:radiation_reaction_1} as follows:
\begin{align}
    \label{eq:radiation_reaction_2}
    f^{\mu}
    & =
        \dfrac{3 e^2_0\dot{x}_{\alpha}(\tau)}{8\pi^2}
        \Big[
            \mathcal{R}
            \int_{-\infty}^{\tau - \Delta}
                \theta(R(\tau,\tau'))
                \dfrac{ h^{\alpha \mu}(\tau,\tau') }
                      { R^{5/2}(\tau,\tau')    }
                \ud \tau'
            +
            \mathcal{R}
            \int_{\tau-\Delta}^{\tau}
                \theta(R(\tau,\tau'))
                \dfrac{ h^{\alpha \mu}(\tau,\tau') }
                      { R^{5/2}(\tau,\tau')        }
                \ud \tau'
        \Big]
\end{align}
Let us focus on the second term, the integration range $\tau' \in (\tau - \Delta, \tau)$.
$\Delta$ is chosen such that $R(\tau,\tau') \neq 0$ for all $\tau - \Delta < \tau' < \tau$.
For most practical purposes, we shall take $\Delta = 1$. In any case, its non-zero value can be absorbed in the 
\emph{renormalized mass}, to be defined below.

In the range $(\tau - \Delta, \tau)$, we can therefore expand both numerator and denominator.
Since $\ve \neq 0$, 
and both $h^{\alpha \mu} (\tau, \tau')$ and $R(\tau,\tau')$ are $O(h^2)$,
we can define the following functions:
\begin{align}
    \label{eq:def_k_function}
    h^{\alpha \mu}(\tau, \tau')
    & \equiv
        h^2 \times k^{\alpha \mu}(\tau, \tau')
    \\
    \label{eq:def_T_function}
    R(\tau, \tau') 
    & \equiv
        h^2 \times T(\tau, \tau')
\end{align}
 
We then find, for the second term in \eqref{eq:radiation_reaction_2}:
\begin{align}
    \label{eq:radiation_reaction_3}
    f_{2}^{\mu}
    & =
        \dfrac{3 e^2_0\dot{x}_{\alpha}(\tau)}{8\pi^2}
        \mathcal{R}
        \int_{\tau-\Delta}^{\tau}
            \dfrac{ h^{\alpha \mu}(\tau,\tau') }
                  { R_{+}^{5/2}(\tau,\tau')    }
            \ud \tau'
    =
    \nonumber
    \\
    & = 
        \dfrac{3 e^2_0\dot{x}_{\alpha}(\tau)}{8\pi^2}
        \mathcal{R}
        \int_{0}^{\Delta}
            \dfrac{ h^2 \times k^{\alpha \mu}(\tau,\tau') }
                  { (h^{2})^{5/2} T_{+}^{5/2}(\tau,\tau')  }
            \ud h
    = 
    \nonumber
    \\
    & = 
        \dfrac{3 e^2_0\dot{x}_{\alpha}(\tau)}{8\pi^2}
        \mathcal{R}
        \int_{0}^{\Delta}
            \dfrac{ k^{\alpha \mu}(\tau,\tau') }
                  { T_{+}^{5/2}(\tau,\tau')  }
            \dfrac{\ud h}{h^3}
    \equiv
        \dfrac{3 e^2_0\dot{x}_{\alpha}(\tau)}{8\pi^2}
        \mathcal{R}
        \int_{0}^{\Delta}
            \dfrac{ \phi^{\alpha \mu}(\tau, h) }
                  { h^{3}                      }
            \ud h
\end{align}
where we have changed integration variable to $\tau' \to h$, 
and defined the function $\phi^{\alpha \mu}(\tau, h) \equiv k^{\alpha \mu}(\tau, \tau-h) / T^{5/2}(\tau,\tau- h)$.

If we consider $\phi^{\alpha \mu}(\tau,h)$ as a \emph{test function}, and $1/h^{3}$, as
a \emph{generalized function}, we can apply Gel'fand regularization \cite{Gelfand1964_1}.
Gel'fand showed that the generalized function $h_{+}^{\lambda}$ 
\begin{align}
    \label{eq:h_function_definition}
    h_{+}^{\lambda}
    & = 
        \begin{cases}
            h^{\lambda} \qquad & h > 0
            \\
            0                  & h < 0
        \end{cases}
\end{align}
has \emph{simple poles} for all negative integers $\lambda = -1, -2, \ldots$.
The \emph{residue} at these poles is given by
\begin{align}
    \label{eq:h_function_residues}
    \stackrel{\text{Res}}{\scriptstyle \lambda = -n}
    h^{\lambda}
    & = 
        \dfrac{(-1)^{n-1}}{(n-1)!}  
        \delta^{(n-1)}(h)
\end{align}

Returning to the full equation of motion
\begin{align}
    \label{eq:radiation_reaction_4}
    M \ddot{x}^{\mu}
    & = 
        \dfrac{3 e^2_0\dot{x}_{\alpha}(\tau)}{8\pi^2}
        \cdot
        \Big[
            \mathcal{R}
            \int_{-\infty}^{\tau - \Delta}
                \dfrac{ h^{\alpha \mu}(\tau,\tau') }
                      { R_{+}^{5/2}(\tau,\tau')    }
                \ud \tau'
            +
            \int_{0}^{\Delta}
                \dfrac{ k^{\alpha \mu}(\tau,\tau') }
                      { T_{+}^{5/2}(\tau,\tau')  }
                \dfrac{\ud h}{h^3}
        \Big]
\end{align}
If the first integral \eqref{eq:radiation_reaction_4} is \emph{regularizable}, then it is \emph{finite}.
The second integral, though non-regularizable, is \emph{renormalizable}.
The renormalization procedure works by dividing both sides by $\Gamma(-3 + \epsilon)$,
and taking the limit of $\epsilon \to 0$.
However, in order to ensure that the equation of motion still maintains a mass term, 
we shall \emph{renormalize the mass as well}. Let us then define
\begin{align}
    \label{eq:renormalized_mass}
    \mmr& = 
        \lim\limits_{\epsilon \to 0} \dfrac{M}{\Gamma(-3 + \epsilon)} 
\end{align}
If $M$ is finite, then $\mmr \to 0$. 
We therefore assume that $m \to \infty$ such that $\mmr \neq 0$, as is often the case in renormalization in quantum field theory,
i.e., in the above-mass-shell case, it is the \emph{renormalized mass} that enters into the dynamics.

Recall that $\ve$ depends on the ratio of $m/M$; 
since $\ve$ is finite, and determined by the dynamics, 
it follows that $m$ is also renormalized. 
This implies further that the quantity $p^{\mu} -e_0 a^{\mu}$ is essentially renormalized in its scale, 
but $\dot{x}^{\mu}$ remains covariantly and finitely determined by the dynamics. 
In the stable situations that we describe below, when $\ve \to 0$, $\smmr \rightarrow \mmr$, 
and it is this value of the mass parameter that may be identified with the Galilean target mass. 

On the other hand, the first integral \eqref{eq:radiation_reaction_4}, 
presumably regularizable, vanishes under the renormalization,
and thus, we are left with the renormalized mass and the \emph{residue} of the second integral, 
given by:
\begin{align}
    \label{eq:radiation_reaction_renormalized_1}
    \mmr 
    \ddot{x}^{\mu}
    & = 
        \dfrac{3 e^2_0\dot{x}_{\alpha}(\tau)}{8\pi^2}
        \dfrac{\ud^2 \phi^{\alpha \mu}(\tau, h)}
              {\ud h^2                         }
        \Big|_{h = 0}
\end{align}

We therefore, need to expand the components of $\phi^{\alpha \mu}$ up to $h^{2}$.
Expansion reveals:
\begin{align}
    \label{eq:k_expansion_h2}
    k^{\alpha\beta}(\tau, \tau - h)
    & = 
        b_{0}^{\alpha \beta}
        + 
        h b_{1}^{\alpha \beta}
        +
        \dfrac{1}{2!}
        h^2 b_{2}^{\alpha \beta}
        + 
        O(h^3)
\end{align}
where:
\begin{align}
    \label{eq:k_expansion_terms}
    \begin{split}
        b_{0}^{\alpha \beta}
        & = 
            \ddot{x}^{\alpha} \dot{x}^{\beta }
            -
            \ddot{x}^{\beta } \dot{x}^{\alpha}
        \\
        b_{1}^{\alpha \beta}
        & = 
            \dfrac{4}{3}
            \Big[
                \dot{x}^{\alpha} \dddot{x}^{\beta } 
                - 
                \dot{x}^{\beta } \dddot{x}^{\alpha}
            \Big]
        \\
        b_{2}^{\alpha \beta}
        & = 
            \dfrac{3}{2}
            \Big[
                \ddddot{x}^{\alpha} \dot{x}^{\beta } 
                - 
                \ddddot{x}^{\beta } \dot{x}^{\alpha} 
            \Big]
            +
            \Big[
                \dddot{x}^{\alpha} \ddot{x}^{\beta } 
                - 
                \dddot{x}^{\beta } \ddot{x}^{\alpha} 
            \Big]
    \end{split}
\end{align}
Similarly, for $T(\tau, \tau - h)$ we find:
\begin{align}
    \label{eq:T_expansion_h2}
    T(\tau, \tau - h)
    & = 
        r_{0}
        + 
        h r_{1}
        +
        \dfrac{1}{2!}
        h^2 r_{2}
        + 
        O(h^3)
\end{align}
and its expansion terms:
\begin{align}
    \label{eq:T_expansion_terms}
    \begin{split}
        r_0
        & = 
            - \dot{x}_{\alpha} \dot{x}^{\alpha} 
        = 
            \ve
        \\
        r_1
        & = 
            \dot{x}_{\alpha} \ddot{x}^{\alpha}
        =
            - \dfrac{1}{2} \dot{\ve}
        \\
        r_2
        & =
            - 
            \dfrac{3}{2} \dot{x}_{\alpha} \dddot{x}^{\alpha} 
            - 
            \dfrac{1}{2} \ddot{x}_{\alpha} \ddot{x}^{\alpha}
    \end{split}
\end{align}

Using the \texttt{Axiom} \emph{computer algebra system} \cite{axiom_1992}, 
the second derivative of $\phi^{\alpha \beta}(\tau, h)$ at $h=0$ is found to be
\begin{align}
    \label{eq:second_derivative_phi}
    \dfrac{\ud^2 \phi^{\alpha \beta}(\tau,h)}
          {\ud h^2                          }
    & = 
        \dfrac
        %% numerator %%
        {
            - 
            10 
            b_{0}^{\alpha \beta} r_{0} r_{2}
            +
            35
            b_{0}^{\alpha \beta} r_{1}^2
            -
            20 
            b_{1}^{\alpha \beta} r_{0} r_{1}
            +
            4 
            b_{2}^{\alpha \beta} r_{0}^2
        }
        %% denominator %%
        {
            4 r_{0}^{9/2}
        }
\end{align}

Defining 
\begin{align}
    \label{eq:renormalized_mass_coefficient}
    D
    & = 
        \dfrac{\mmr \cdot 8\pi^2}{3 e_0^2 }
\end{align}
we finally obtain the renormalized Lorentz force equation:
\begin{align}
    \label{eq:renormalized_lorentz_force_1}
    \begin{split}
        \mr 
        \ddot{x}^{\mu}
        & = 
            \dfrac{1}{8 \ve^{9/2}}
            \Bigg[
                \dot{x}^{\mu}
                \Big(
                    5
                    \ve \dot{\ve} 
                    \ddot{\ve} 
                    + 
                    \dfrac{45}{6} 
                    \ve \dot{\ve} 
                    \rho 
                    - 
                    \dfrac{35}{8}
                    \dot{\ve}^3
                    + 
                    2 \ve^2 \dot{x}_{\alpha} \ddddot{x}^{\alpha}
                \Big)
        \\
        & \qquad\qquad 
                +
                \ddot{x}^{\mu}
                \Big(
                    - 
                    4\ve^2 \ddot{\ve} 
                    - 
                    3 \ve^2 \rho 
                    +
                    \dfrac{35}{2} \ve \dot{\ve}^2
                \Big)
                - 
                6 \ve^2 \dot{\ve} \dddot{x}^{\mu}
                + 
                2 \ve^3 \ddddot{x}^{\mu}
            \Bigg]
    \end{split}
\end{align}
where we have defined an additional scalar $\rho$ given by:
\begin{align}
    \label{eq:rho_definition}
    \rho 
    & =
        \ddot{x}_{\alpha} \ddot{x}^{\alpha}
    =
        \ddot{x}_{\mu} \ddot{x}^{\mu}
    ,
\end{align}
where we note that for any derivative higher than unity, the fifth component  of $x^{5}$ is annihilated,
as $x^{5}(\tau) \equiv \tau$ and $\dot{x}^{5}(\tau) \equiv 1$.

This Lorentz force equation is of $4^{\text{th}}$-order.
It is more natural to express the higher order part in terms of lower order.
From equation \eqref{eq:renormalized_lorentz_force_1}, we find that
\begin{align}
    \label{eq:renormalized_lorentz_force_2}
    \begin{split}
        {M^{\mu}}_{\nu}
        \ddddot{x}^{\nu}
        & = 
            - 
            \dfrac{2}{\ve^2}
            \Bigg[
                \dot{x}^{\mu}
                \Big(
                    5
                    \ve \dot{\ve} 
                    \ddot{\ve} 
                    + 
                    \dfrac{45}{6} 
                    \ve \dot{\ve} 
                    \rho 
                    - 
                    \dfrac{35}{8}
                    \dot{\ve}^3
                \Big)
        \\
        & \qquad\qquad 
                +
                \ddot{x}^{\mu}
                \Big(
                    - 
                    4\ve^2 \ddot{\ve} 
                    - 
                    3 \ve^2 \rho 
                    +
                    \dfrac{35}{2} \ve \dot{\ve}^2
                    -
                    \mr 
                    \times 
                    8 
                    \ve^{9/2}
                \Big)
                - 
                6 \ve^2 \dot{\ve} \dddot{x}^{\mu}
            \Bigg]
    \end{split}
\end{align}
where 
\begin{align}
    {M^{\mu}}_{\nu}
    & = 
        \ve 
        {\delta^{\mu}}_{\nu} 
        + 
        \dot{x}^{\mu} \dot{x}_{\nu}
\end{align}

One can eliminate the factor ${M^{\mu}}_{\nu}$ by contracting with $\dot{x}_{\mu}$:
\begin{align}
    \label{eq:M_mu_nu_dot_x_mu}
    {M^{\mu}}_{\nu} \dot{x}_{\mu}
    & = 
        \ve 
        \dot{x}_{\nu}
        + 
        \dot{x}^{\mu} \dot{x}_{\mu} \dot{x}_{\nu}
    =
        \ve \dot{x}_{\nu}
        +
        (- \ve - 1) 
        \dot{x}_{\nu}
    =
        - 
        \dot{x}_{\nu}
\end{align}
Thus:
\begin{align}
    \label{eq:renormalized_lorentz_force_2_step_2}
    \begin{split}
        \dot{x}_{\mu}
        {M^{\mu}}_{\nu}
        \ddddot{x}^{\nu}
        & = 
            - \dot{x}_{\nu} \ddddot{x}^{\nu}
        =
        \\
        & = 
            - 
            \dfrac{2}{\ve^2}
            \Bigg[
                \dot{x}_{\mu}
                \dot{x}^{\mu}
                \Big(
                    5
                    \ve \dot{\ve} 
                    \ddot{\ve} 
                    + 
                    \dfrac{45}{6} 
                    \ve \dot{\ve} 
                    \rho 
                    - 
                    \dfrac{35}{8}
                    \dot{\ve}^3
                \Big)
        \\
        & \qquad\qquad 
                +
                \dot{x}_{\mu}
                \ddot{x}^{\mu}
                \Big(
                    - 
                    4\ve^2 \ddot{\ve} 
                    - 
                    3 \ve^2 \rho 
                    +
                    \dfrac{35}{2} \ve \dot{\ve}^2
                    -
                    \mr 
                    \times 
                    8 
                    \ve^{9/2}
                \Big)
                - 
                6 \ve^2 \dot{\ve} 
                \dot{x}_{\mu}
                \dddot{x}^{\mu}
            \Bigg]
        \\
        & = 
            - 
            \dfrac{2}{\ve^2}
            \Bigg[
                (- \ve - 1)
                \Big(
                    5
                    \ve \dot{\ve} 
                    \ddot{\ve} 
                    + 
                    \dfrac{45}{6} 
                    \ve \dot{\ve} 
                    \rho 
                    - 
                    \dfrac{35}{8}
                    \dot{\ve}^3
                \Big)
        \\
        & \qquad\qquad 
                - 
                \dfrac{1}{2} \dot{\ve}
                \Big(
                    - 
                    4\ve^2 \ddot{\ve} 
                    - 
                    3 \ve^2 \rho 
                    +
                    \dfrac{35}{2} \ve \dot{\ve}^2
                    -
                    \mr 
                    \times 
                    8 
                    \ve^{9/2}
                \Big)
                - 
                6 \ve^2 \dot{\ve} 
                \dot{x}_{\mu}
                \dddot{x}^{\mu}
            \Bigg]
    \end{split}
\end{align}

We then find a simpler equation of motion for $\ddddot{x}^{\mu}$:
\begin{align}
    \label{eq:renormalized_lorentz_force_3}
    \ddddot{x}^{\mu}
    & = 
        \dfrac{2}{\ve^2}
        \Bigg[
            \dot{x}^{\mu} \dot{\ve}
            \Big(
                \dfrac{35}{8} \dot{\ve}^2 
                - 
                4 \mr \ve^{7/2}
            \Big)
            +
            \ddot{x}^{\mu}
            \Big(       
                4\ve \ddot{\ve}
                + 
                3 \ve \rho 
                - 
                \dfrac{35}{2} \dot{\ve}^2
                + 
                8 \mr \ve^{7/2}
            \Big)
            +
            6 \ve \dot{\ve}
            \dddot{x}^{\mu}
        \Bigg]
\end{align}

\subsection{Scalar set of equations}
Equation \eqref{eq:renormalized_lorentz_force_3} is a vector equation of $4^{\text{th}}$ order, 
up to $3^{\text{rd}}$ derivative of $x^{\mu}$. 
Therefore, the of this system is 12D, \emph{if one excludes external fields}, 
comprising the dynamical variables $\dot{x}^{\mu}, \ddot{x}^{\mu}$ and $\dddot{x}^{\mu}$.
If external fields are also present, then $x^{\mu}$ is also a part of configuration-space,
increasing the dimensionality of the system to 16D.

Reduction of phase space is possible by suitable contractions of \eqref{eq:renormalized_lorentz_force_3}.

For this purpose, we define yet an additional scalar, $\eta$,
\begin{align}
    \label{eq:eta_definition}
    \eta & = 
        \dddot{x}_{\mu} \dddot{x}^{\mu}
\end{align}

Using the definitions of $\ve, \rho, \eta$ 
and their derivatives, and
using the following relations
\begin{align*}
    \dot{x}_{\mu} \ddddot{x}^{\mu}  & = - \dfrac{1}{2} \dddot{\ve} - \dfrac{3}{2}\dot{\rho}
    &
    \ddot{x}_{\mu} \ddddot{x}^{\mu} & = \dfrac{1}{2}\ddot{\rho} - \eta
    &
    \dddot{x}_{\mu} \ddddot{x}^{\mu}& = \dfrac{1}{2} \dot{\eta}
\end{align*}
we eventually find the following set of equations:
\begin{align}
    \label{eq:renormalized_scalar_equations_set}
    \begin{split}
        \dddot{\ve}
        & = 
            - 
            3 \dot{\rho} 
            + 
            2 (\ve+1) K_1
            + 
            \dot{\ve} K_2
            +
            (\ddot{\ve} + 2\rho)
            K_3
        \\
        \ddot{\rho}
        & = 
            2 \eta
            -
            \dot{\ve} K_1
            + 
            2 \rho K_2
            + 
            \dot{\rho} K_3
        \\
        \dot{\eta}
        & = 
            - 
            (\ddot{\ve} + 2\rho)K_1
            +
            \dot{\rho}  K_2
            +
            2 \eta K_3
    \end{split}
\end{align}
where we have defined the following auxiliary functions, for clarity:
\begin{align}
    \label{eq:K_functions}
    \begin{split}
        K_1
        & = 
            \dfrac{2\dot{\ve}}{\ve^2}
            \Big[
                \dfrac{35}{8} 
                \dot{\ve}^2
                -
                4 \mr \ve^{7/2}
            \Big]
        \\
        K_2
        & = 
            \dfrac{2}{\ve^2}
            \Big[
                4 \ve \ddot{\ve}
                + 
                3 \ve \rho
                - 
                \dfrac{35}{2} \dot{\ve}^2
                + 
                8 \mr \ve^{7/2}
            \Big]
        \\
        K_3
        & = 
            12 \dfrac{\dot{\ve}}{\ve}
    \end{split}
\end{align}

Equations \eqref{eq:renormalized_scalar_equations_set} 
indicate that the reduced phase space includes the scalars $\{\ve, \dot{\ve}, \ddot{\ve}, \rho, \dot{\rho}, \eta\}$,
all of them containing up to $3^{\text{rd}}$ order derivatives in $x^{\mu}$.
Numerical study of these equations is given in the following section \ref{sec:numerical_simulation}.

The $K_i$ functions \eqref{eq:K_functions} can be interpreted as \emph{potential scalars}, as they practically 
form the potential landscape of the system.

Before that, we can still study two simple solutions analytically, both of them, for \emph{constant mass-shell deviation $\ve$}.

\subsection{The motions with constant mass-shell deviation}
In this section, we are seeking solutions such that $\ve$ remains constant.
Thus, this immediately leads to:
\begin{align*}
    \dot{\ve} & = \ddot{\ve} = \dddot{\ve} = 0
\end{align*}
In this case, only $K_2 = 6 \rho / \ve + 16 D \ve^{3/2}$ remains non-zero, and the only non-trivial 
equation is the one for $\ddot{\rho}$:
\begin{align*}
    \ddot{\rho} & = 2 \eta + 2\rho \Big( 6 \dfrac{\rho}{\ve} + 16 D \ve^{3/2} \Big)
\end{align*}
However, from the equation for $\dddot{\ve}= 0$, we immediately find $\dot{\rho} = 0$, 
and therefore, we are left with:
\begin{align}
    \label{eq:renormalized_scalar_equations_set_constant_epsilon}
    \begin{split}
        \dot{\ve} & = \ddot{\ve } = \dddot{\ve} = \dot{\rho} = \dot{\eta} = 0
        \\
        \eta 
        & = 
            - \rho 
            \Big(
                \dfrac{6\rho}{\ve} 
                + 
                16 D\ve^{3/2}
            \Big)
    \end{split}
\end{align}
In this case, all $\ve, \rho$ and $\eta$ remain constant.

Clearly, the simplest case would be to set $\rho = \eta = 0$, which is the case of \emph{uniform motion}.
This is understandable, as uniform motion should not cause radiation reaction.

The other case would be a non-zero $\rho, \eta$, which means that even for accelerated particle, 
a particular orbit exists for which $\ve$ remains constant.

Both cases, as seen below, are actually unstable, and the case of non-zero $\rho, \eta$
is unstable numerically as well, eventually leading the 
particle either back to mass-shell $\ve \to 0$ or to infinite mass $\ve \to \infty$.

\section{Numerical simulation and results}
\label{sec:numerical_simulation}
In this section, numerical simulation of the set of equations \eqref{eq:renormalized_scalar_equations_set} is shown.
The scalar system comprises the $6$ scalars $\ve, \dot{\ve}, \ddot{\ve}, \rho, \dot{\rho}$ and $\eta$.
The scalar system offers a reduced phase-space description of the system, 
and its divergence is completely unrelated to a particular Lorentz system.
However it lacks the coordinate representation offered in the original vector system 
given in \eqref{eq:renormalized_lorentz_force_3}.
Therefore, when it aids clarity, this vector equation set was also used to produce numerical results.

\subsection{Numerical methods}
The system was studied both in the scalar set of equations \eqref{eq:renormalized_scalar_equations_set} 
and the original vector one \eqref{eq:renormalized_lorentz_force_3}.
The simulation was written in C++, employing a standard RK4
integrator \footnote{Runge-Kutta method of $4^{\text{th}}$ order with $5^{\text{th}}$ order error estimatation.}
taken from \emph{GSL} \cite{gsl_2009}, and rewritten to use high-precision arithmetic provided by 
\emph{MPFR} \cite{Fousse:2007:MMP} and \emph{MPFRCPP} \cite{mpfrcpp_2011} as its higher level interface.
Results are plotted mostly with \emph{matplotlib} \cite{Hunter:2007}, and some with \emph{VisIt} \cite{Childs:2005:ACS}.

All the numerical results of this paper were obtained with free and open-source software.

\subsection{Results}
\begin{figure}
    \begin{center}
        \includegraphics[width=12cm]{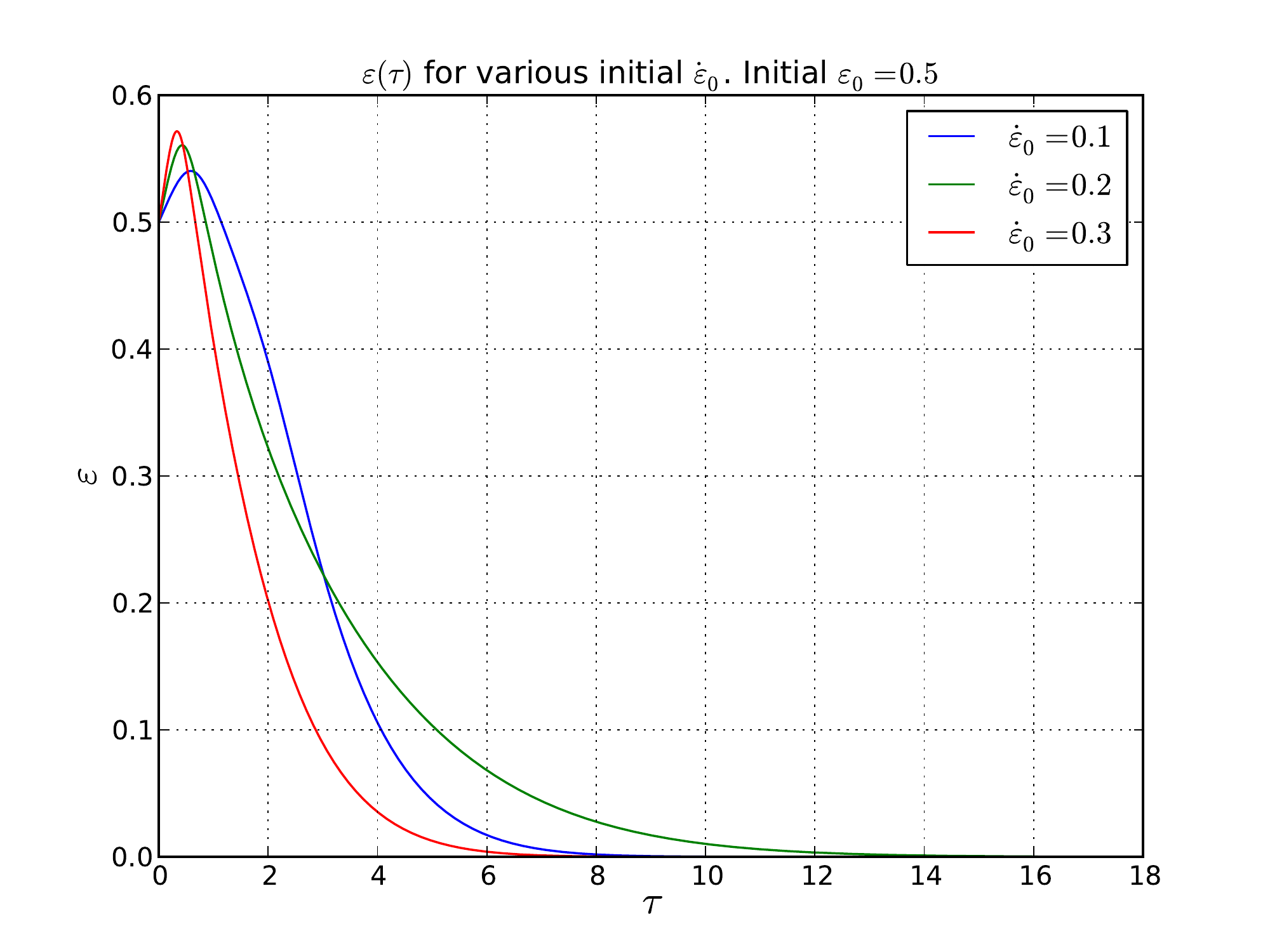}
    \end{center}
    \caption{An example of $3$ solutions of \eqref{eq:renormalized_scalar_equations_set}, where the mass-shell deviation scalar 
             $\ve(\tau)$ starting from non-zero value and eventually 
             falling smoothly towards the on-shell $\ve \to 0$, asymptotically approaching uniform motion.}
    \label{fig:plotone_epsilon_falls_to_zero}
\end{figure}

Let us begin with the simpler cases, where $\ve$ smoothly goes to $0$.
In figure \ref{fig:plotone_epsilon_falls_to_zero}, $3$ plots of $\ve(\tau)$ for $3$ initial conditions for $\dot{\ve}_0$ are shown.
The other variables, namely $\ddot{\ve}_0, \rho_0, \dot{\rho}_0$ and $\eta_0$ all begin at $0$.
Even though $\dot{\ve}$ tends to increase $\ve$, eventually, $\ve$, with all derivatives, smoothly go to zero.

\begin{figure}
    \begin{center}
        \includegraphics[width=14cm]{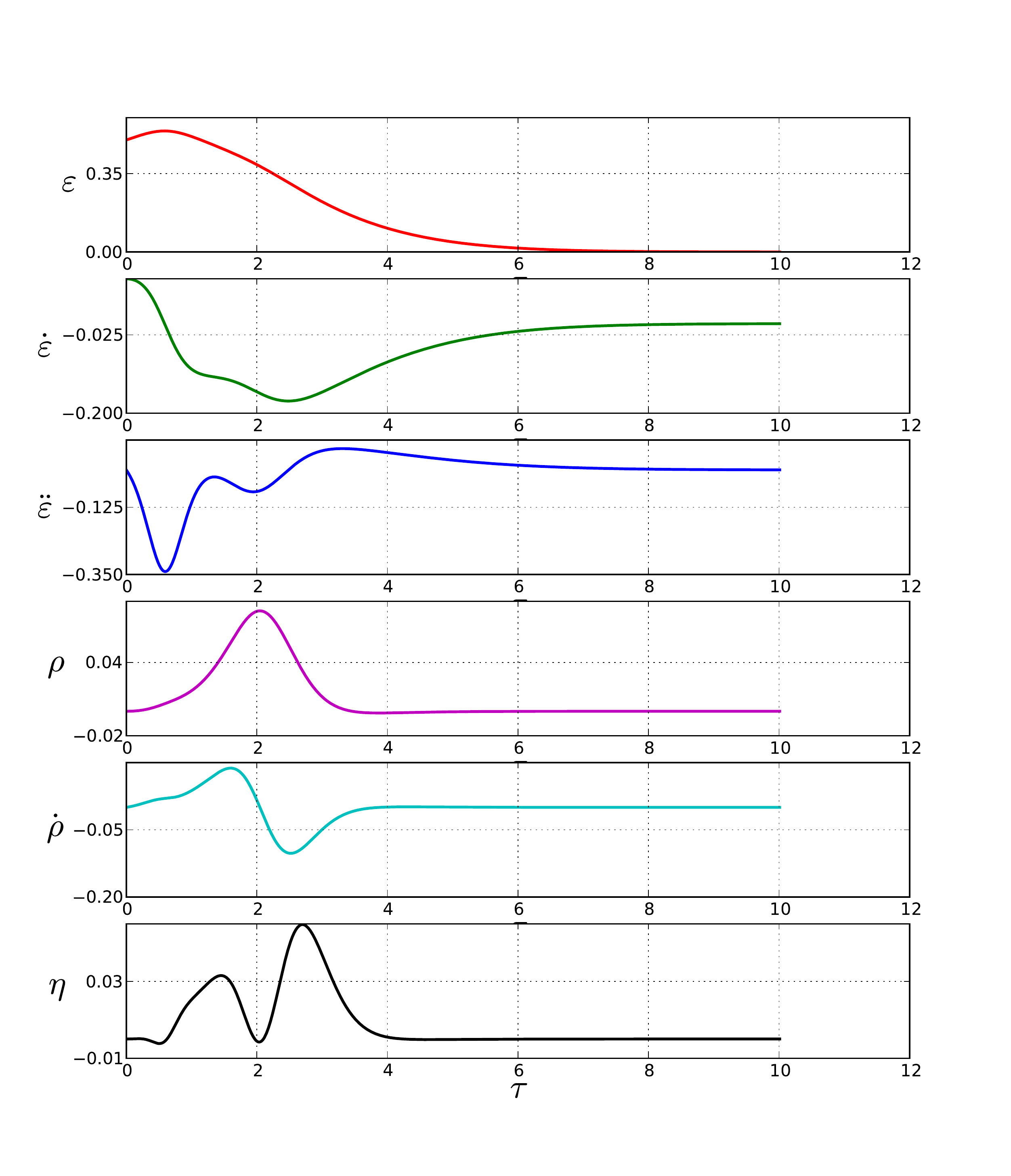}
    \end{center}
    \caption{A converging $\ve \to 0$ example, where the full set of scalar variables is shown, with initial 
             conditions $\ve_0 = 0.5, \dot{\ve}_0 = 0.1, \ddot{\ve}_0 = \rho_0 = \dot{\rho}_0 = \eta_0 = 0$.}
    \label{fig:plottwo_full_scalar_system_falls_to_zero}
\end{figure}

In figure \ref{fig:plottwo_full_scalar_system_falls_to_zero}, the entire set of scalar variables
showing the entire behavior of the system, as it smoothly falls down towards the mass shell 
and asymptotically into unifom motion.

We note that the asymptotic limit of $\ve=0$ cannot be handled in this set, as the point $\ve=0$ is not defined for this system.
Therefore, in order to handle the \emph{transition} between above-mass-shell, on-mass-shell and below-mass-shell behavior, one
has to develop and solve \eqref{eq:radiation_reaction_1} for the other cases as well.
This would be studied in a succeeding publication.

\begin{figure}
    \begin{center}
        \includegraphics[width=14cm]{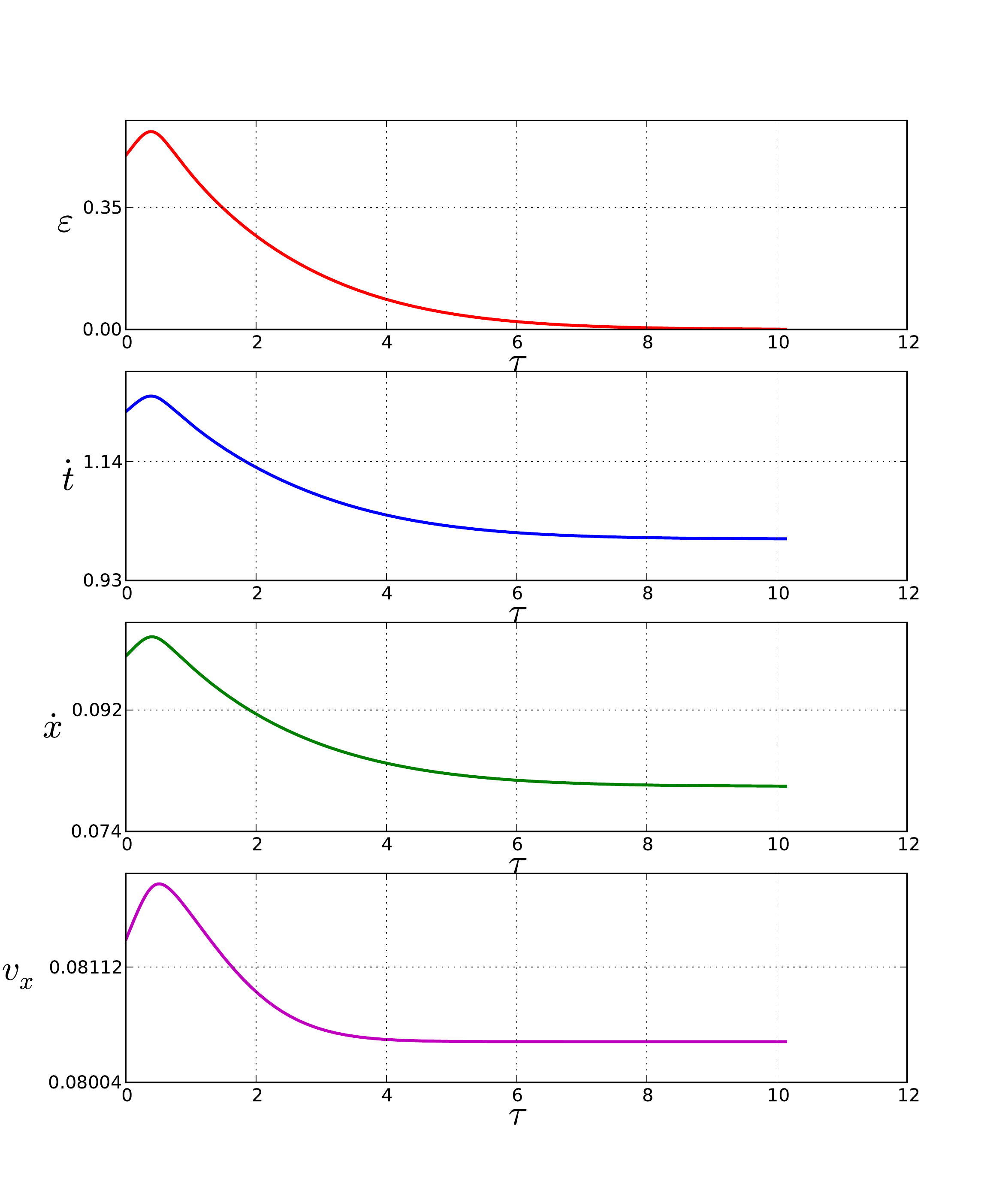}
    \end{center}
    \caption{
                A converging $\ve \to 0$ example, using the vector system of equations \eqref{eq:renormalized_lorentz_force_3}.
                Initial conditions are given in vector form for $\dot{x}^{\mu}_0, \ddot{x}^{\mu}_0$, and where $\dddot{x}^{\mu}_0 = 0$.
                The entire motion is in the $t-x$ plane, and clearly, both $\dot{x}$ and $\dot{t}$ converge towards uniform motion.
            }
    \label{fig:plotthree_vector_equation_epsilon_falling_down_to_zero}
\end{figure}

In figure \ref{fig:plotthree_vector_equation_epsilon_falling_down_to_zero}, a similar system is solved using 
the vector system of equations \eqref{eq:renormalized_lorentz_force_3}, 
with initial conditions in $\dot{x}^{\mu}, \ddot{x}^{\mu}$ and $\dddot{x}^{\mu}$.
It is evident that as $\ve \to 0$, both $\dot{t}(\tau)$ and $\dot{x}(\tau)$ approach a constant value, 
and as a consequence, $v_x = \dot{x} / \dot{t} = \ud x / \ud t$  goes to a finite limit as well (less than one).

\begin{figure}
    \begin{center}
        \includegraphics[width=14cm]{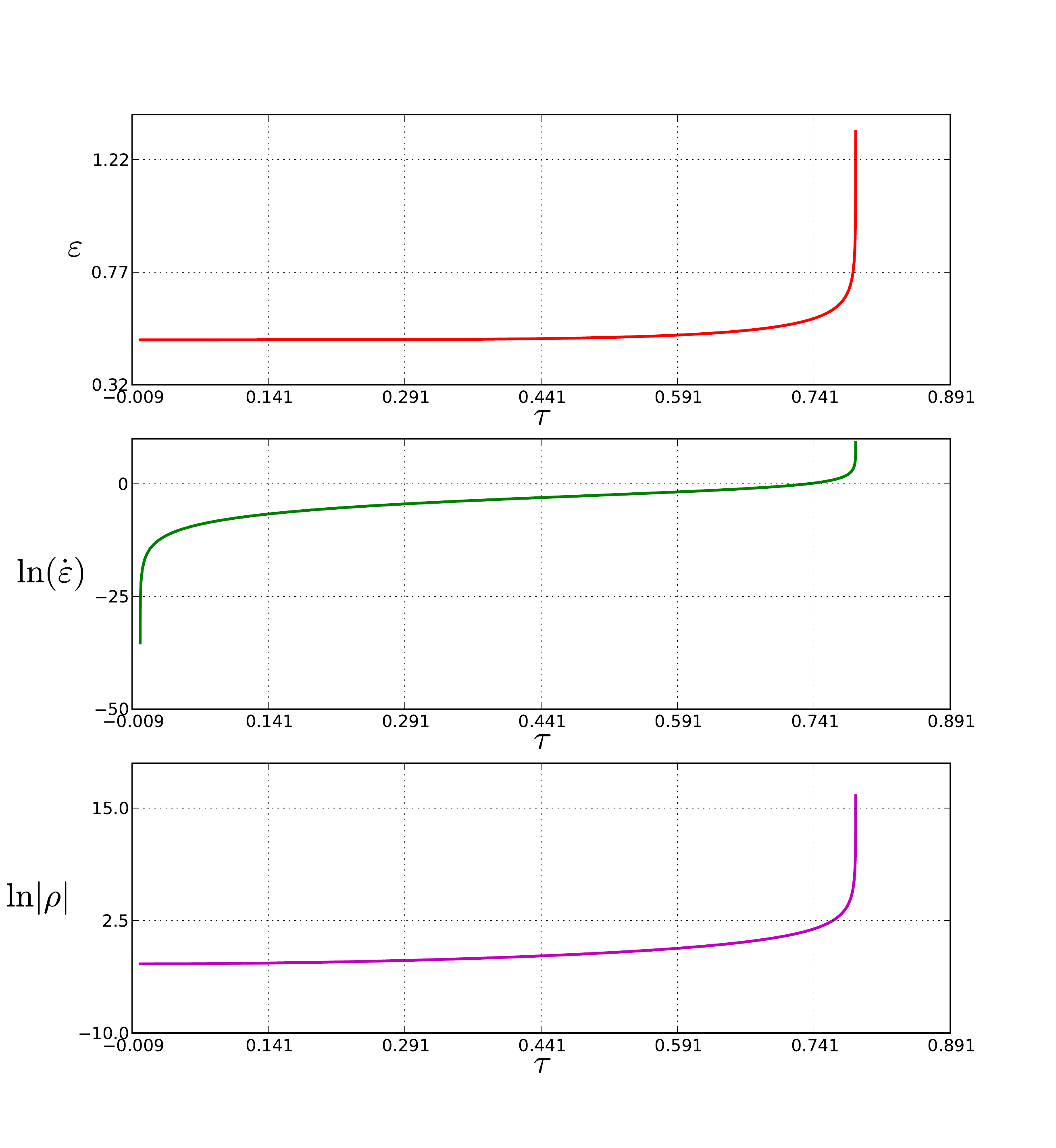}
    \end{center}
    \caption{
                A diverging solution example $\ve \to \infty$, solved using the scalar system of equations 
                \eqref{eq:renormalized_scalar_equations_set}.
                Initial conditions are $\ve_0 = 0.5, \rho_0 = -0.1$ and all the rest are $0$.
                It strongly diverges at a finite $\tau \approx  0.787$. Only $\ve, \ln(\dot{\ve})$ and $\ln|\rho|$ 
                are shown, though the behavior of the other terms are very similar.
                $\ve, \dot{\ve}$ and $\ddot{\ve}$ are positive, whereas $\rho, \dot{\rho}, \eta$ are negative.
            }
    \label{fig:plotfour_scalar_equation_epsilon_diverges}
\end{figure}

In figure \ref{fig:plotfour_scalar_equation_epsilon_diverges}, a strongly divergent solution of the scalar system
is shown. In this case, $\rho$ is \emph{negative}, whereas in ordinary SR dynamics, we find, for an on-mass-shell 
particle, that $\ddot{x}_{\mu} \ddot{x}^{\mu} = \vb{a}_{\text{lf}}^2 > 0$, where $\vb{a}_{\text{lf}}$ 
is the acceleration observed in a \emph{co-moving frame}.
In this particular case shown in \ref{fig:plotfour_scalar_equation_epsilon_diverges}, 
the initial $\rho_0$ was already negative, and therefore, cannot be associated
with a simple 3-vector acceleration $\vb{a}_{\text{lf}}$ in a co-moving frame.
It is possible, however, to arrive at a diverging case, even when initially, $\dot{\rho}_0 > 0$,
but in all cases of the diverging system, eventually, $\rho \to -\infty$, 
which is extreme non-SR behavior.

\begin{figure}
    \begin{center}
        \includegraphics[width=13cm]{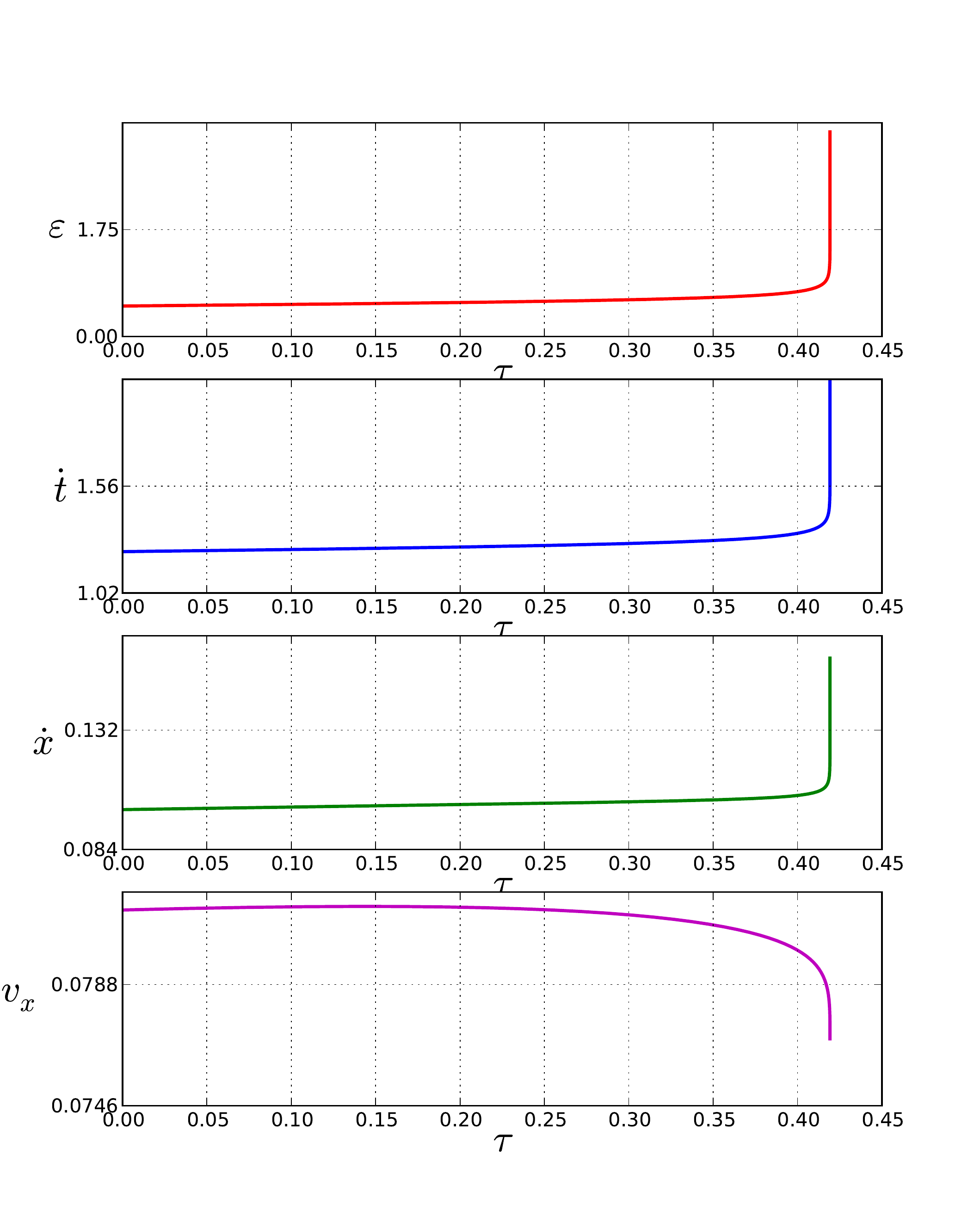}
    \end{center}
    \caption{
                A diverging solution example $\ve \to \infty$, 
                formulated under the vector equation system given in \eqref{eq:renormalized_lorentz_force_3}.
                Only the plots for $\ve, \dot{t}, \dot{x}$ and $v_x = \dot{x} / \dot{t}$ are shown.
                Interestingly, even though $\dot{x}$ and $\dot{t}$ diverge, the 3-velocity $v_x$ converges
                to a lower and finite value (less than unity).
            }
    \label{fig:plotsix_vector_equation_epsilon_diverges}
\end{figure}

In figure \ref{fig:plotsix_vector_equation_epsilon_diverges}, another strongly divergent solution of the \emph{vector system}
is shown (see \eqref{eq:renormalized_lorentz_force_3}).
In spite of the divergence, the 3-velocity $v_x$ plotted seems to strongly converge towards a finite value.

\begin{figure}
    \begin{center}
        \includegraphics[width=14cm]{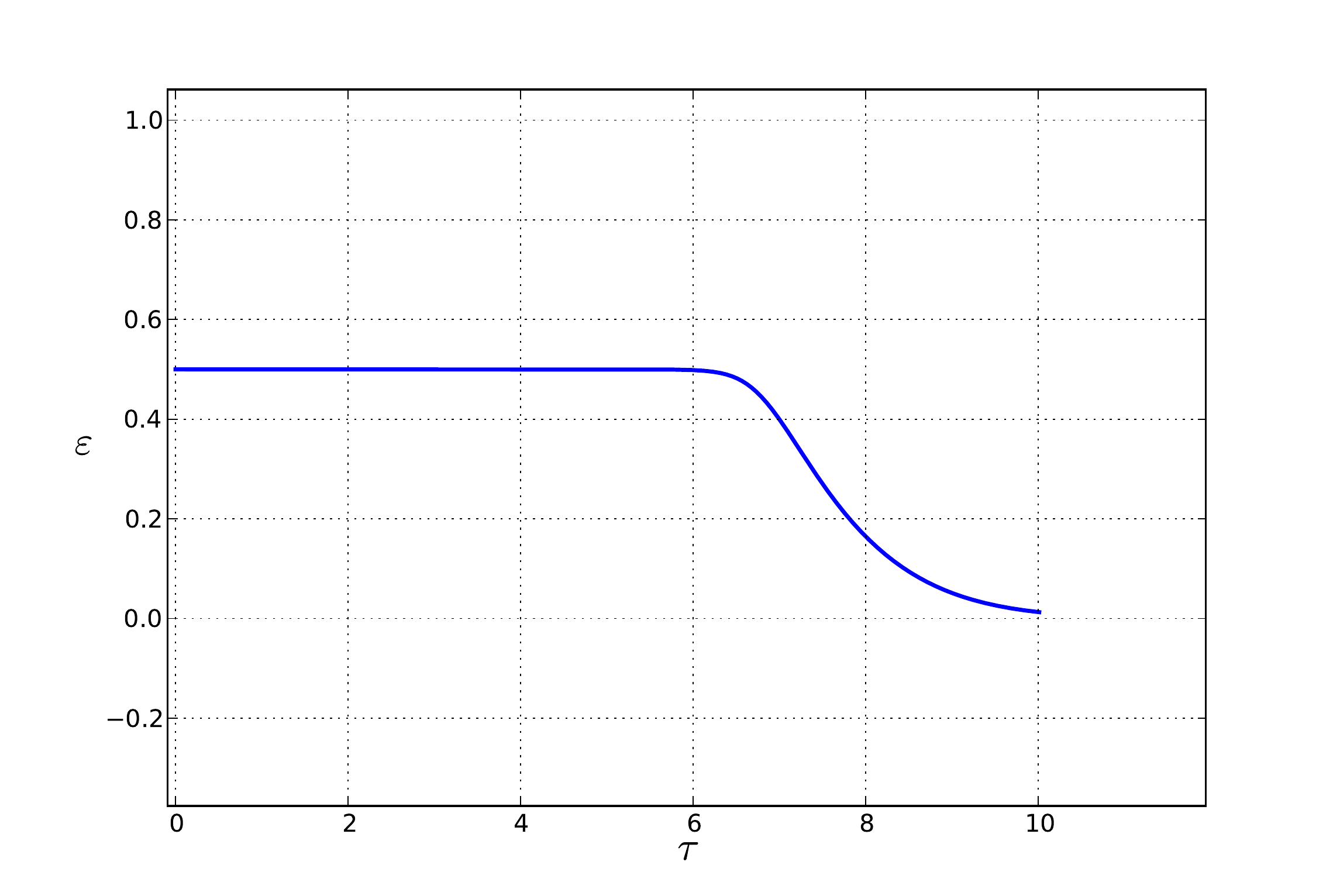}
    \end{center}
    \caption{
                A case of initially constant $\ve$ in an accelerated system $\rho_0 = 0.1, \eta_0 \approx -0.685$,
                where, eventually, $\ve \to 0$ possibly due to the finite, though large, numerical precision.
                The constant non-zero value of $\ve$ appears to be unstable.
            }
    \label{fig:plotseven_scalar_equation_epsilon_constant_but_eventually_converges_to_zero}
\end{figure}

And finally, in figure \ref{fig:plotseven_scalar_equation_epsilon_constant_but_eventually_converges_to_zero},
a case for an \emph{almost} constant $\ve = 0.5$ case is shown. 
Eventually, the system decays due to the finite precision; it is therefore  unstable.

\subsection{Analysis}
The converging cases depicted in figures 
(
\ref{fig:plotone_epsilon_falls_to_zero} ,
\ref{fig:plottwo_full_scalar_system_falls_to_zero} ,
\ref{fig:plotthree_vector_equation_epsilon_falling_down_to_zero}
)
show, eventually, smooth uniform \emph{on-mass-shell} motion.
In particular, $\rho$, which is, essentially, the $3$-acceleration squared $\vb{a}_{\text{lf}}^2$ as seen in a co-moving frame, 
is always positive, and eventually, falls to $\rho \to 0$, indicating uniform motion.

The diverging case $\ve \to \infty$ requires a more detailed investigation.
$\ve \gg 0$ indicates that $|\dot{t}| \gg |\vb{\dot{x}}|$, and therefore
% % % % and since
% % % % \begin{align}
% % % %     \label{eq:dot_t_vs_dot_x}
% % % %     \dot{t}
% % % %     & = 
% % % %         \dfrac{E - e_0 a^{0} }{M}
% % % %     &
% % % %     \vb{\dot{x}}
% % % %     & = 
% % % %         \dfrac{\vb{p} - e_0 \vb{a}}{M}
% % % % \end{align}
it follows that $\dot{t}^2 - \dot{x}^2 = \smmr^2 / \mmr^2 \gg 1$.

% % % % % % % Using the 3-velocity
% % % % % % % \begin{align}
% % % % % % %     \label{eq:v_dot_x}
% % % % % % %     \vb{v}
% % % % % % %     & = 
% % % % % % %         \dfrac{\ud \vb{x}}{\ud t}
% % % % % % %     = 
% % % % % % %         \dfrac{\vb{\dot{x}}}{\dot{t}}
% % % % % % % \end{align}
% % % % % % % we could deduce that $\vb{v} \to 0$, i.e., that the 3-velocity actually observed in a laboratory,
% % % % % % % would converge towards zero. However, this is a non-Lorentz-covariant behavior, since if we move to a different
% % % % % % % moving laboratory, a different 3-velocity would be measured.

However
%%Using \eqref{eq:mass_shell_definition_x_dot_x_dot}, we can find
\begin{align}
    \ve
    & = 
        \dot{t}^2 
        - 
        \vb{\dot{x}}^2 
        -
        1
    =
        \dot{t}^2 
        (1 - \vb{v}^2)
        -
        1
    \gg
        0
\end{align}
and therefore, $\vb{v}$ must be less than unity.
% % % % Therefore, the factor $1 - \vb{v}^2 = \gamma^{-2}$ where $\gamma$ is the standard dilation factor in SR,
% % % % is not necessarily $1$, and $\vb{v}$ could approach any value less than one.

At the present we do not have a way to deduce the actual limiting value of $\vb{v}$ in a diverging scenario, as 
the numerical simulation cannot continue the integration of a system where $\ve \to \infty$ indefinitely,
though it does indicate a decreasing bound on it.

It seems then, that \emph{even in the diverging $\ve \to \infty$ } case, the 3-velocity converges towards a finite value,
which would manifest itself as \emph{uniform motion} in any frame, i.e., similar to the $\ve \to 0$ case.

Thus, the theory is well defined even for extreme unstable off-shell deviation.

\subsection{Origin of divergence}
\begin{figure}
    \begin{center}
        \includegraphics[width=14cm,page=1]{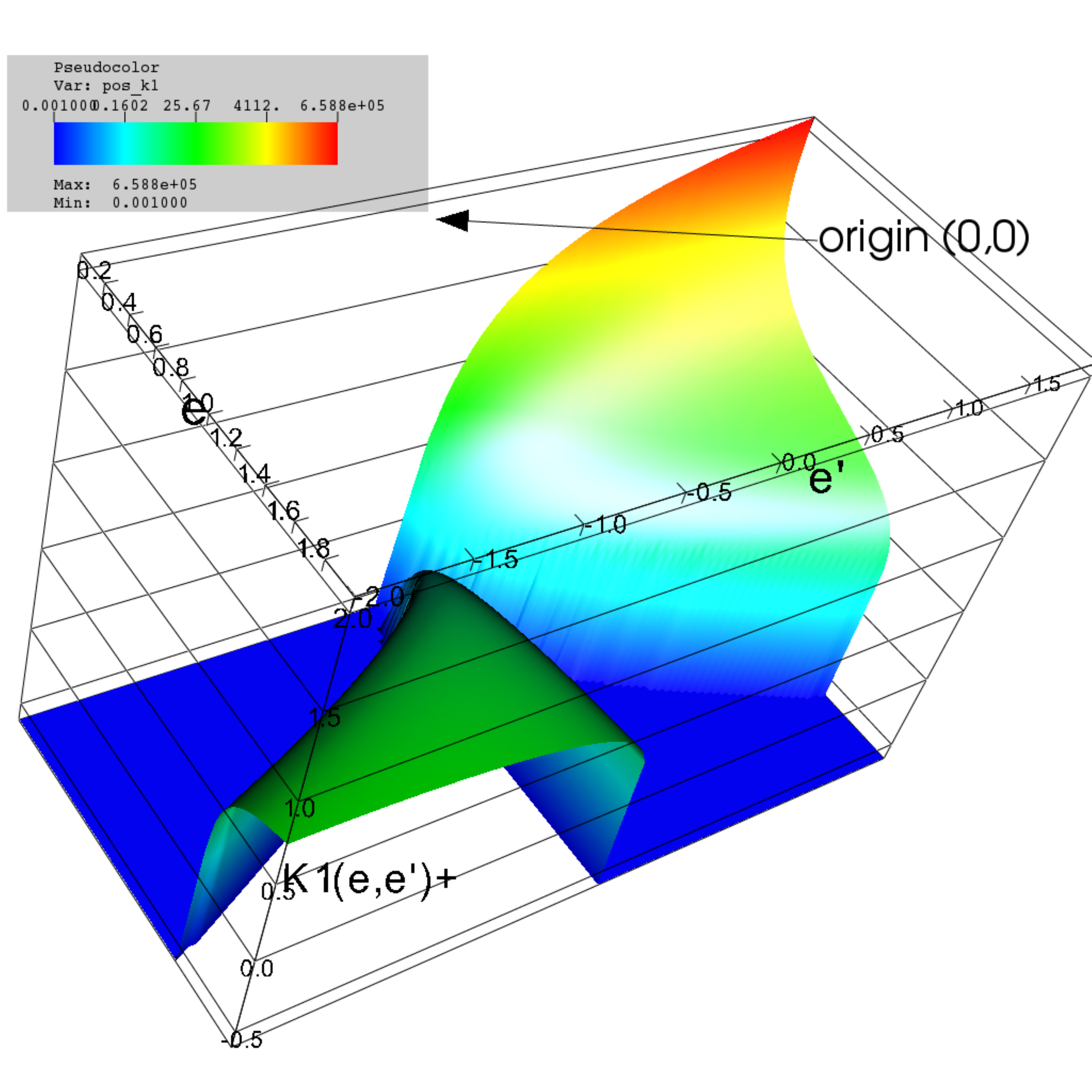}
    \end{center}
    \caption{
                Logarithmically colored elevated plot of $K_1^{+}(\ve,\dot{\ve})$ (labeled as $\mathtt{e}$ and $\mathtt{e'}$ 
                in this \emph{VisIt} \cite{Childs:2005:ACS} plot), 
                where $K_{i}^{+}$ refers to the \emph{positive} part of $K_i$, the negative part being zero.
            }
    \label{fig:k1_plus_plot}
\end{figure}

\begin{figure}
    \begin{center}
        \includegraphics[width=14cm,page=2]{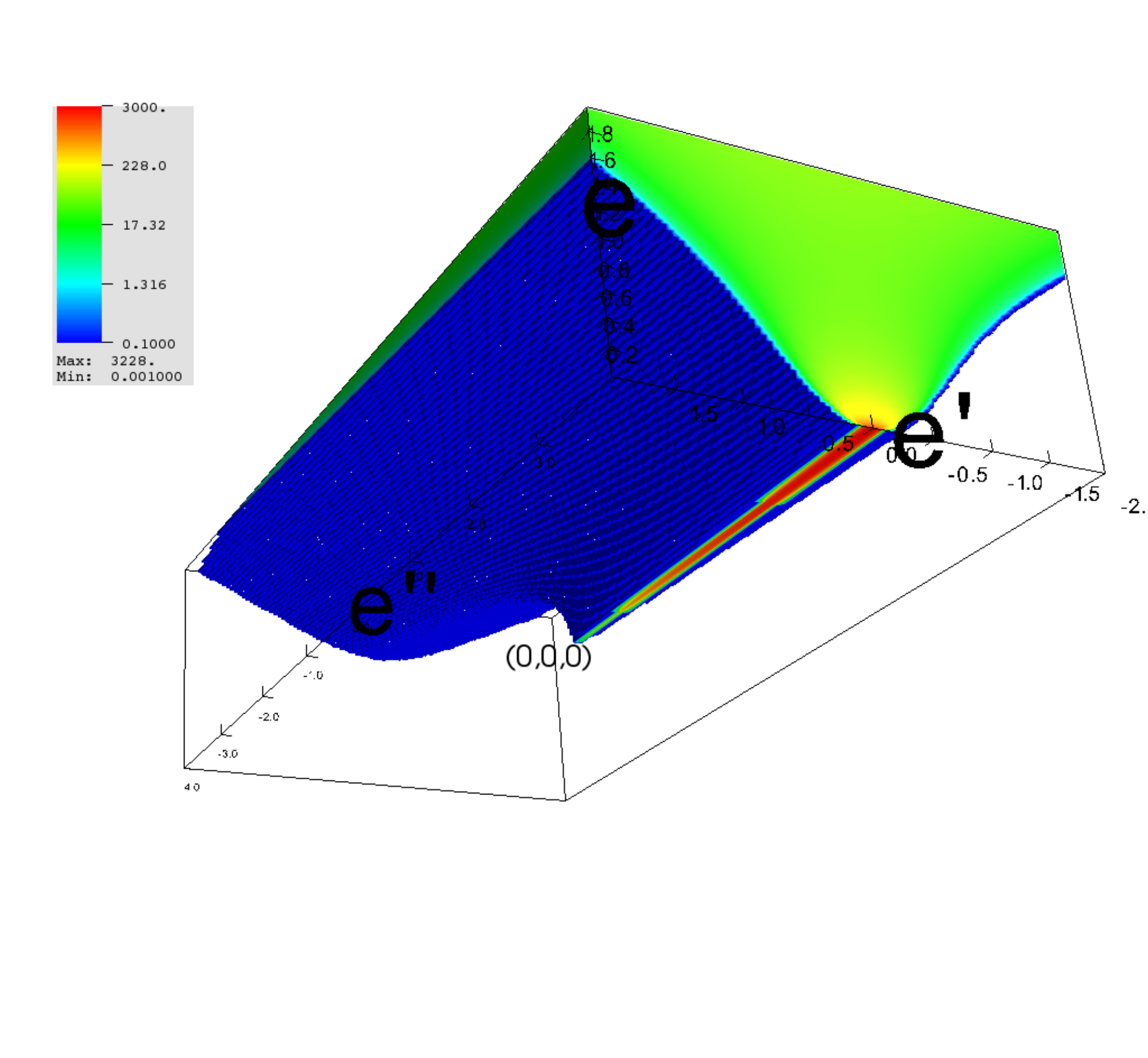}
    \end{center}
    \caption{
                Logarithmically colored plot of $K_2^{+}(\ve,\dot{\ve},\ddot{\ve}, \bullet)$ 
                on a rectangular mesh
                ($\ve, \dot{\ve}$ and $\ddot{\ve}$ are labelled as $\mathtt{e}, \mathtt{e'}$ and $\mathtt{e''}$ 
                in this \emph{VisIt} \cite{Childs:2005:ACS} plot).
                The $0$ valued nodes of the mesh were excluded to reveal the positive ones.
                Note that though the plot is 3D and shows only the dependence on $\ve, \dot{\ve}$ and $\ddot{\ve}$,
                the dependence on the extra $\rho$ is co-linear with $\ddot{\ve}$, and therefore, 
                there's no loss of information in this reduced function plot.
                The origin $(\ve, \dot{\ve}, \ddot{\ve}) = (0,0,0)$ is marked in the middle of the plot.
            }
    \label{fig:k2_plus_plot}
\end{figure}

\begin{figure}
    \begin{center}
        \includegraphics[width=14cm,page=3]{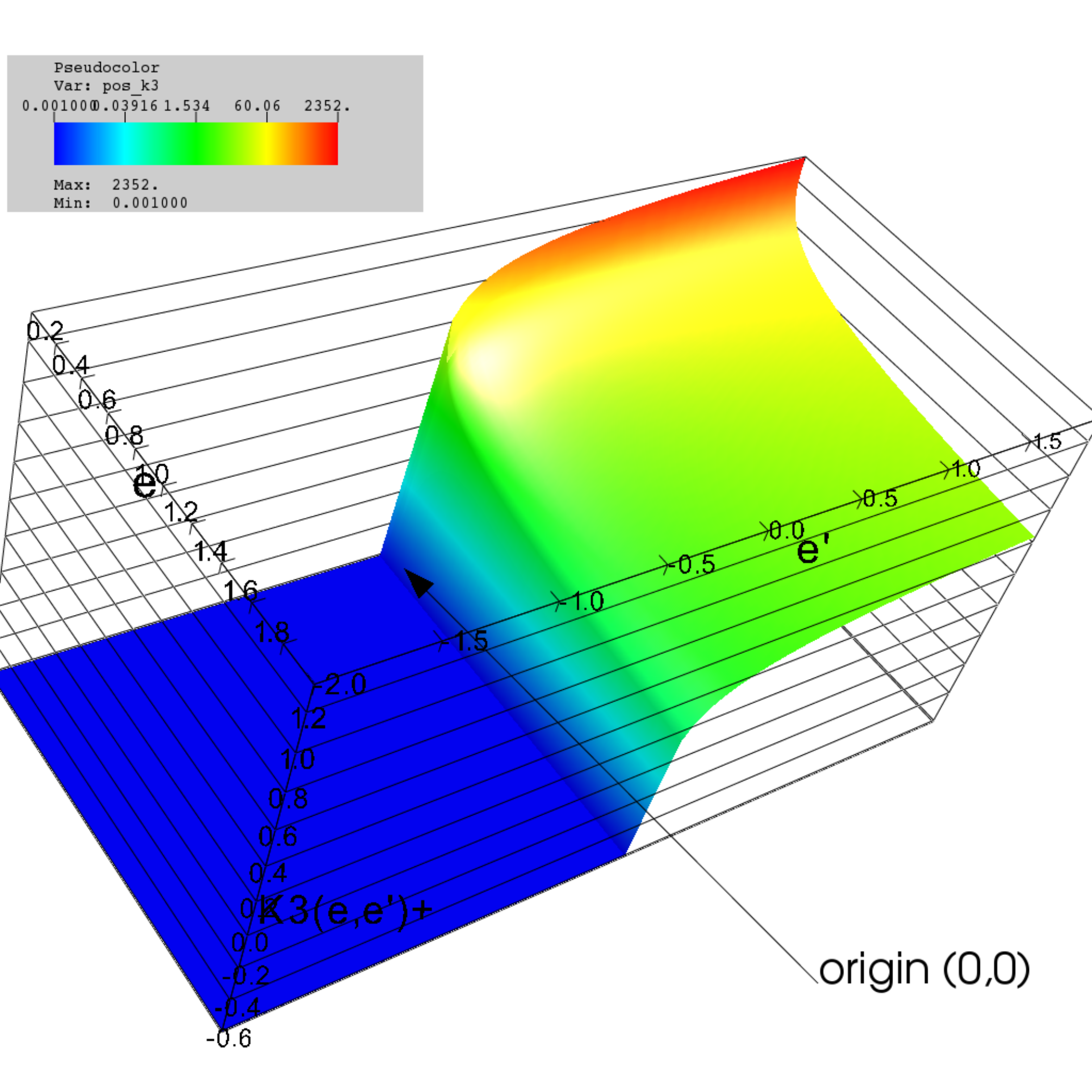}
    \end{center}
    \caption{
                Logarithmically colored elevated plot of $K_3^{+}(\ve,\dot{\ve})$ 
                (labeled as $\mathtt{e}$ and $\mathtt{e'}$ in this \emph{VisIt} \cite{Childs:2005:ACS} plot).
                Note that the region $\dot{\ve} < 0$ is entirely zero, as $K_3(\ve, \dot{\ve}) < 0$ there.
            }
    \label{fig:k3_plus_plot}
\end{figure}

The set of scalar equations \eqref{eq:renormalized_scalar_equations_set} seems as a linear combination
of the $K_i$ functions, coupled linearly to the dynamic variables.
Therefore, it seems useful to study the behavior of the $K_i$ functions, as possible origin of instability.
Logarithmically colored plots of the \emph{positive} parts of $K_i$ functions are shown in figures 
\ref{fig:k1_plus_plot}, \ref{fig:k2_plus_plot} and \ref{fig:k3_plus_plot},
where we defined the notation:
\begin{align}
    \label{eq:K_i_plus_definition}
    K_{i}^{+} 
    & = 
        \theta( K_i ) K_i 
    =
    \begin{cases}
            K_{i}       \qquad  &       K_i \geq 0 
            \\
            0                   &       K_i < 0
        \end{cases}
\end{align}

Of these functions, only $K_2$ has substantial \emph{positive} value even near the $(\ve,\dot{\ve}) \to (0,0)$ 
origin\footnote{The exact point $\ve = 0$ is outside the domain of the above-shell analysis presented in this paper.},
as both $K_1$ and $K_3$ have a factor of $\dot{\ve}$.
This can be seen as $K_2$ value depends linearly on the value of $4\ddot{\ve} + 3\rho$, as well, and thus, even when $\dot{\ve} = 0$,
it may possess a positive value.

The positive values of $K_i$ normally push the system towards $\dot{\ve} > 0$, i.e., towards the 
\emph{diverging} $\ve \to \infty$. 

To see this, we may begin with the equation for $\dddot{\ve}$ in \eqref{eq:renormalized_scalar_equations_set}.
The diverging case is characterized by $\ve \to \infty$, and therefore, $\dot{\ve}, \ddot{\ve}, \dddot{\ve} \to \infty$ as well.
In the equation for $\dddot{\ve}$, we find that this divergence is indeed maintained, as long as $\dot{\rho} \to - \infty$, 
as well as $K_i \to \infty$, and finally, the coefficient of $K_3$ has to be positive, leading to $|\rho| < \ddot{\ve}/2$.
I.e., though in this case $\rho \to - \infty$, its divergence is controlled by $\rho > - \ddot{\ve}/2$.

This characterization of divergence is also consistent with the equations of motion for $\ddot{\rho}$ and $\dot{\eta}$,
as long as $\eta \to - \infty$ for the diverging case.

Summarizing, the conditions of instability are:
\begin{align}
    \label{eq:conditions_of_instability}
    \begin{split}
        K_i                             & \to + \infty
        \\
        \ve, \dot{\ve}, \ddot{\ve}      & \to + \infty
        \\
        \rho, \dot{\rho}. \eta          & \to - \infty
    \end{split}
\end{align}

\begin{figure}
    \begin{center}
        \includegraphics[width=14cm]{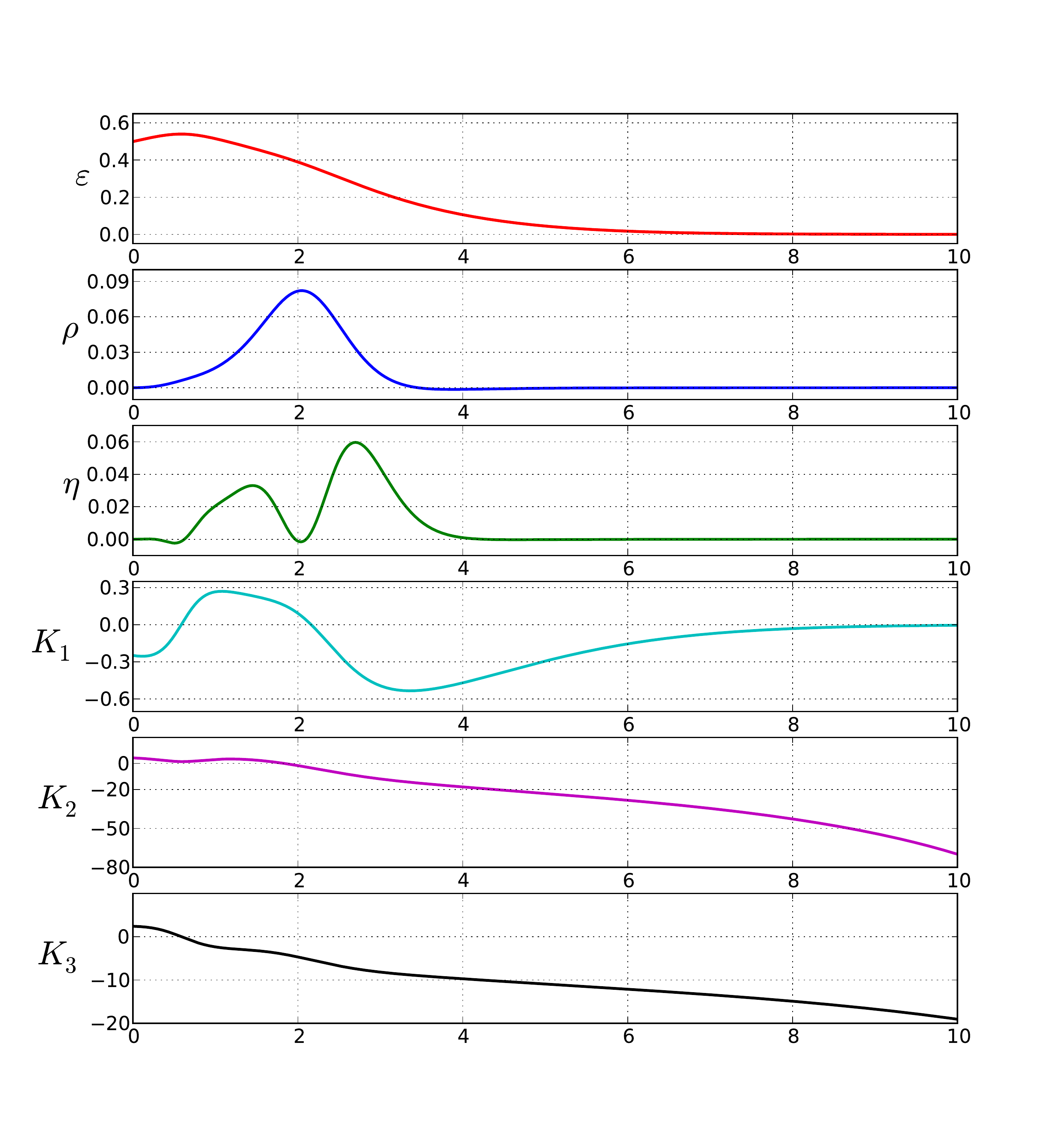}
    \end{center}
    \caption{
                A case of a converging scalar system \eqref{eq:renormalized_scalar_equations_set}, 
                where the value of the \emph{scalar potentials $K_i$} are also shown.
                Clearly, all potential scalars $K_i \leq 0$ as $\tau \to \infty$,
                which leads to system stability.
            }
    \label{fig:converging_k1_k2_k3}
\end{figure}

\begin{figure}
    \begin{center}
        \includegraphics[width=13cm]{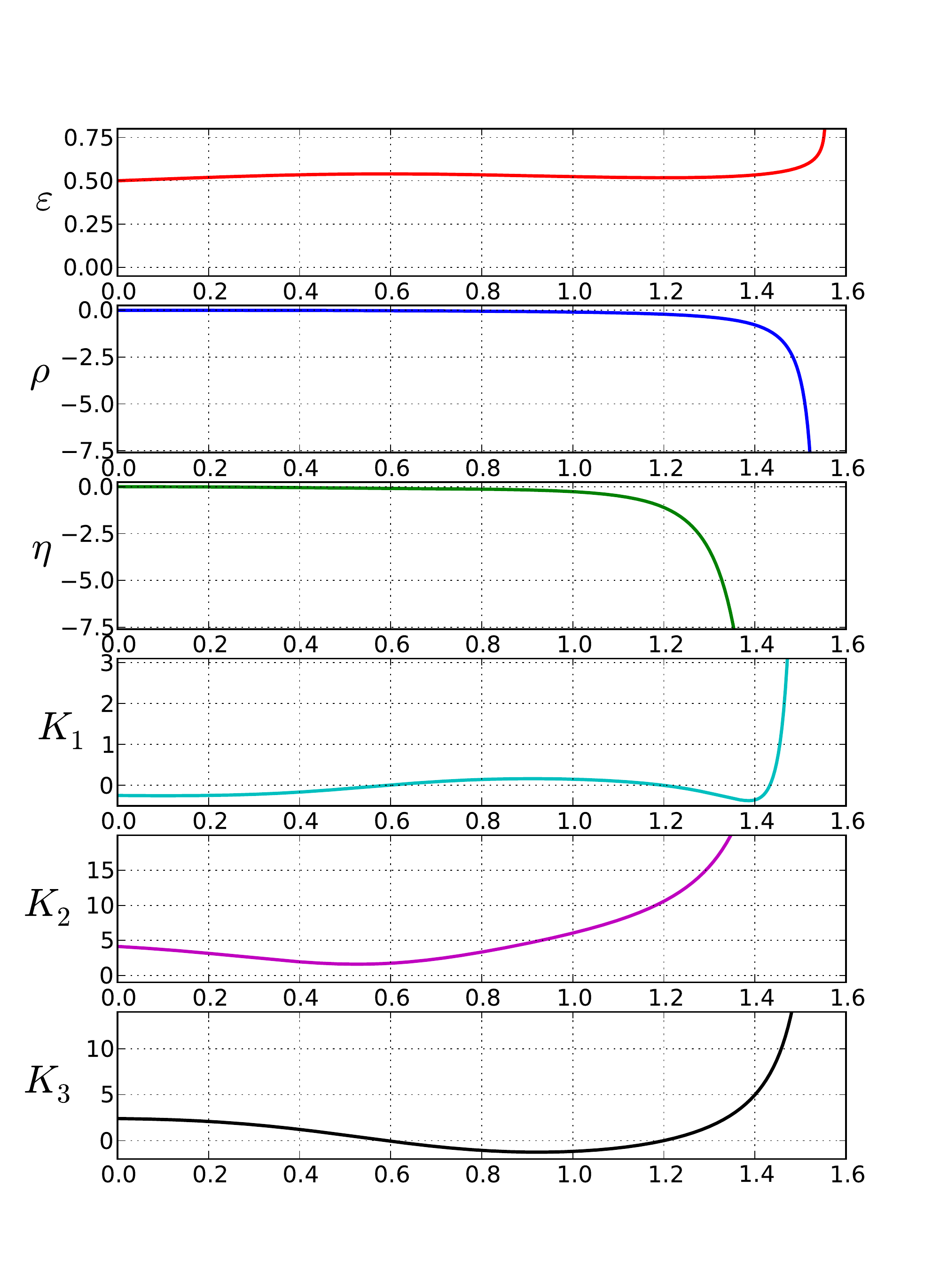}
    \end{center}
    \caption{
                A case of a \emph{diverging} scalar system \eqref{eq:renormalized_scalar_equations_set},
                where the value of the \emph{scalar potentials $K_i$} are also shown.
                Clearly, all potential scalars $K_i \to \infty$ as $\tau \to 1.6$,
                in which the system becomes unstable.
            }
    \label{fig:diverging_k1_k2_k3}
\end{figure}

In figure \ref{fig:converging_k1_k2_k3}, the value of the potential scalars $K_i$ is shown 
for a \emph{converging case} $\ve \to 0$. As $\tau$ increases, eventually, $K_1 \to 0, K_2, K_3 \to -\infty$, 
or, $K_i \leq 0$.
On the other hand, figure \ref{fig:diverging_k1_k2_k3}, a diverging system is shown where all the scalars 
$K_i \to \infty$, and the entire system is attracted towards instability.

\subsection{Stability analysis}
Another way to see the divergence/convergence, is to use linearized stability analysis.
The scalar equation set \eqref{eq:renormalized_scalar_equations_set} can be cast 
into a formal first order system
\begin{align}
        \label{eq:renormalized_scalar_equations_set_dynamical_system_approach}
        \dot{x}
        & = 
            f(x)
\end{align}
where $x \in \mathbb{R}^{6}$:
\begin{align}
        x^{T} & = ( \ve, \dot{\ve}, \ddot{\ve}, \rho, \dot{\rho}, \eta )
\end{align}
and $f(x)$ is a 6-vector defined as:
\begin{align}
    \label{eq:f_vector_of_scalar_system}
    f(x)
    & = 
        \begin{pmatrix}
            \dot{\ve}
            \\
            \ddot{\ve}
            \\
            -3 \dot{\rho} + 2(\ve + 1)           K_1 + \dot{\ve} K_2 + (\ddot{\ve} + 2\rho) K_3
            \\
            \dot{\rho}
            \\
            2 \eta        - \dot{\ve}            K_1 + 2 \rho    K_2 + \dot{\rho}           K_3
            \\
                          - (\ddot{\ve} + 2\rho) K_1 + \dot{\rho}K_2 + 2\eta                K_3
        \end{pmatrix}
\end{align}

The Jacobian $J(x)$ is defined as:
\begin{align}
    \label{eq:def_jacobian}
    {J^{a}}_{b} 
    & = 
        \dfrac{\partial f^{a} (x) }
              {\partial x^{b}     }
\end{align}
On each point $x$ along the orbit, ${J^{a}}_{b}$ determines the \emph{local} behavior,
in the sense that for each eigenvalue $\lambda_i$ with its associated eigenvector $v_i$, 
determine the \emph{local} behavior of the motion.

\begin{figure}
    \begin{center}
        \includegraphics[width=14cm]{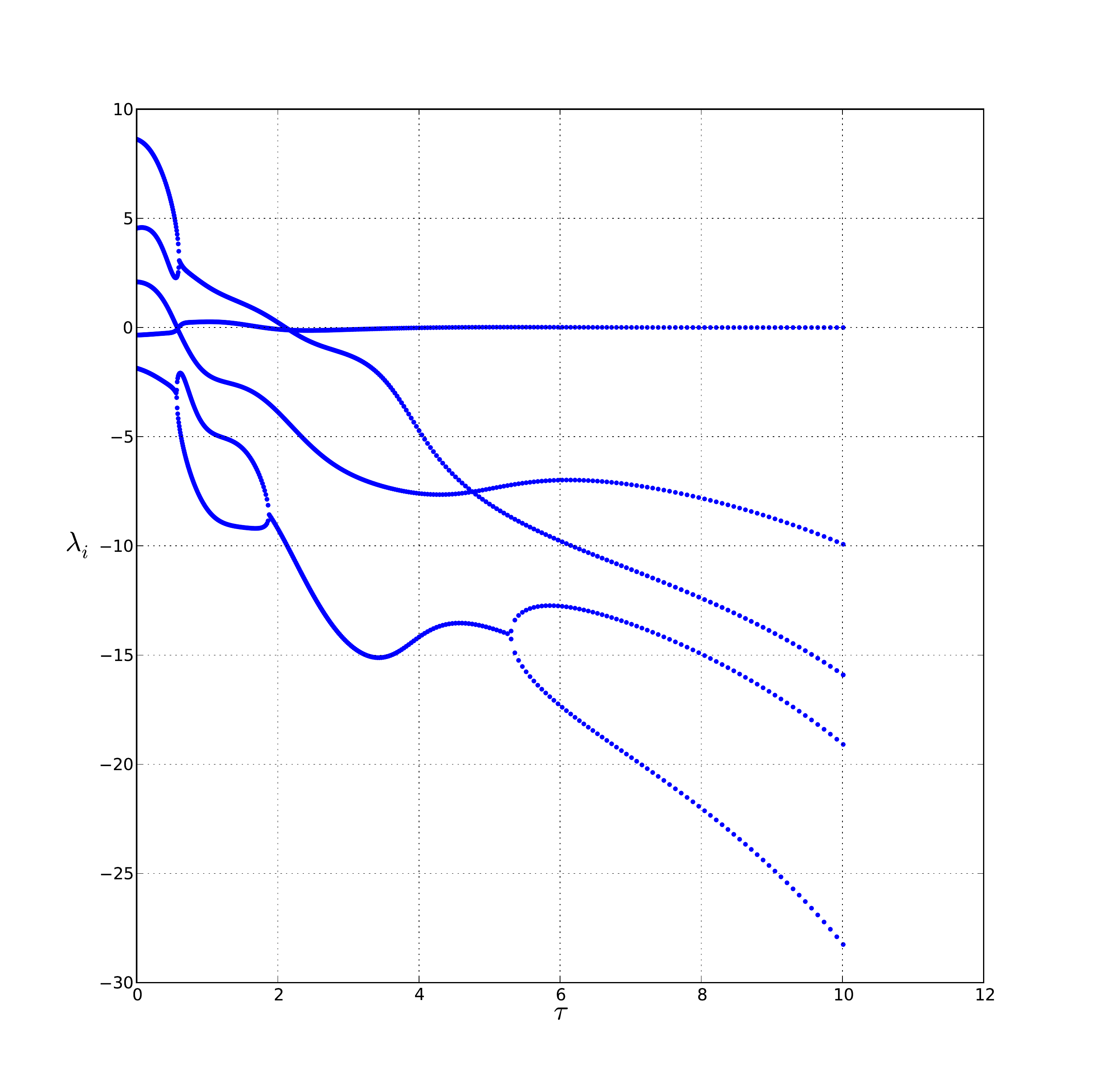}
    \end{center}
    \caption{
                Plot of the $6$ eigenvalues of ${J^{a}}_{b}$ of \eqref{eq:def_jacobian} along the orbit 
                of the scalar system \eqref{eq:renormalized_scalar_equations_set},
                in a case of convergence towards zero mass-shell deviation.
                Aside from one eigenvalue, all tend towards stable negative values as $\tau \to \infty$.
                Only the \emph{real part} is shown.
            }
    \label{fig:eigenvalues_of_converging_system}
\end{figure}

\begin{figure}
    \begin{center}
        \includegraphics[width=14cm]{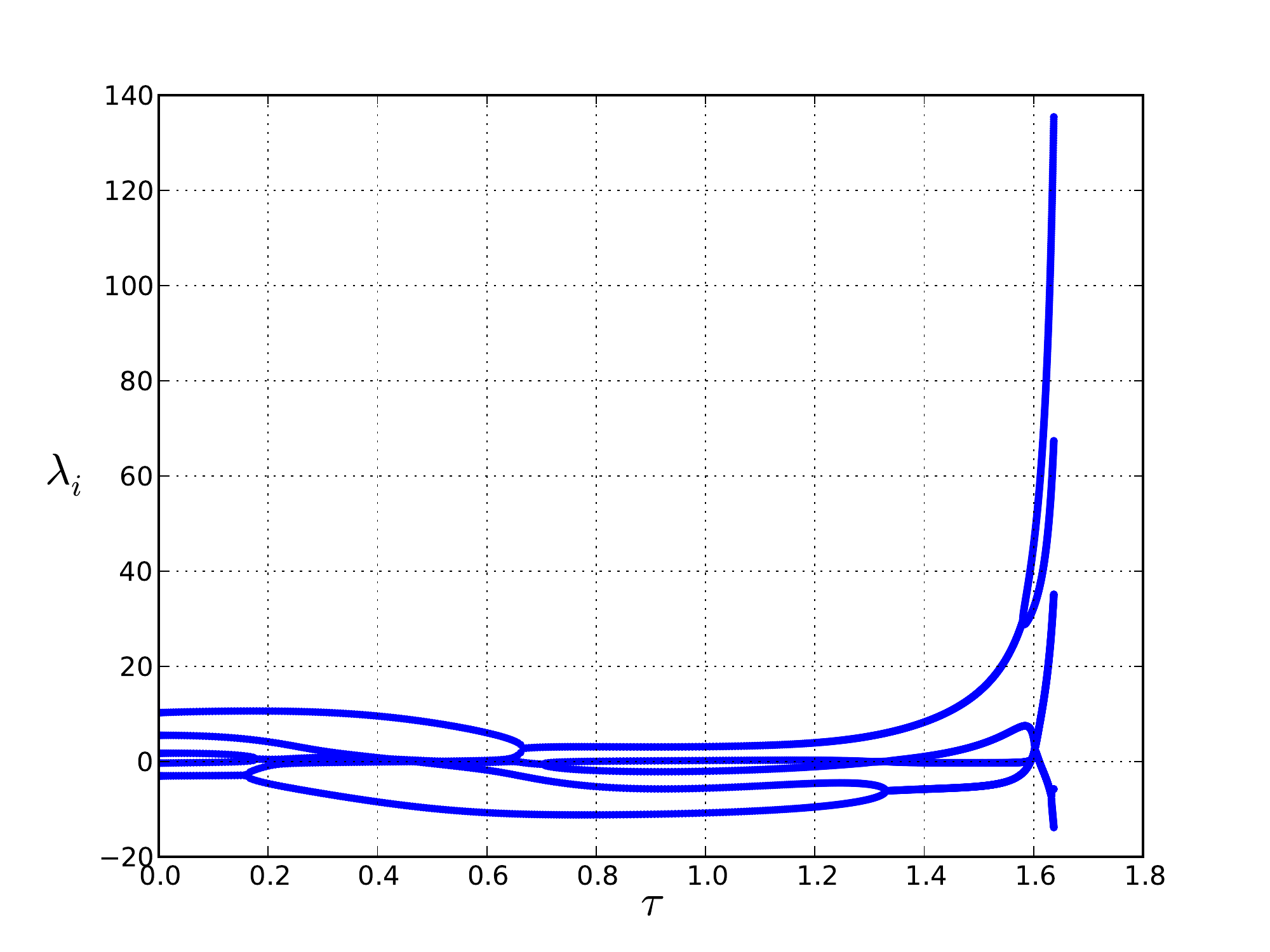}
    \end{center}
    \caption{
                Plot of the $6$ eigenvalues of ${J^{a}}_{b}$ of \eqref{eq:def_jacobian} 
                along the orbit of a \emph{diverging} scalar system \eqref{eq:renormalized_scalar_equations_set}.
                $4$  eigenvalues go towards $+\infty$, one goes towards $-\infty$, and one towards $0$.
            }
    \label{fig:eigenvalues_of_diverging_system}
\end{figure}

In figure \ref{fig:eigenvalues_of_converging_system}, the real part of the eigenvalues of \eqref{eq:def_jacobian} 
along the orbit of a \emph{converging system} are shown. 
In figure \ref{fig:eigenvalues_of_diverging_system}, on the other hand, the same plots for the case of a \emph{diverging system} are shown. 

This suggests that the origin of the 6D phase space $\ve, \dot{\ve}, \ldots, \eta \to 0$ is a stable stationary point,
as eigenvalues near it are almost all negative. On the other hand, stationary points not near the 6D origin, 
are \emph{unstable}.

\begin{figure}
    \begin{center}
        \includegraphics[width=14cm]{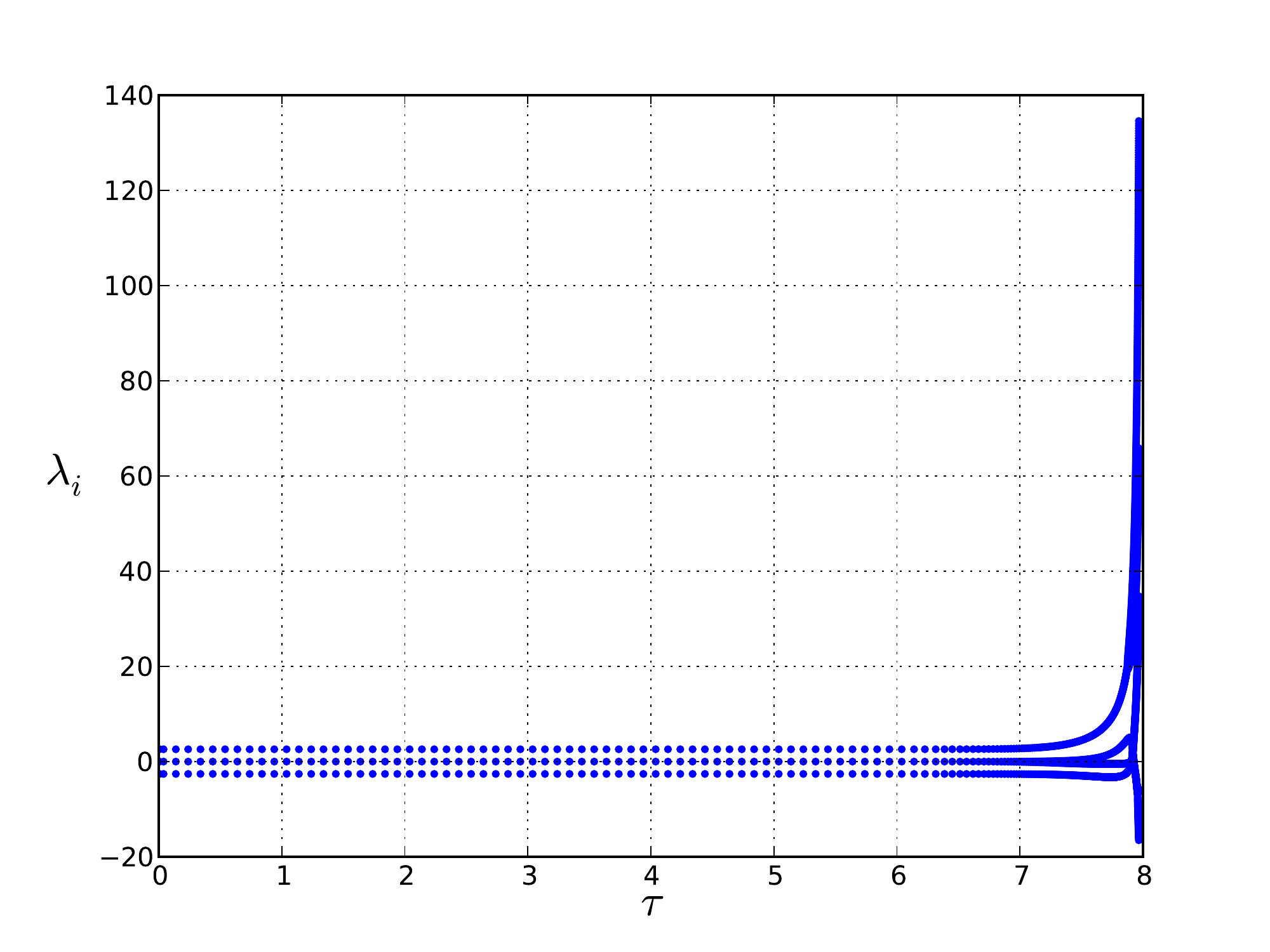}
    \end{center}
    \caption{
                Plot of the $6$ eigenvalues of ${J^{a}}_{b}$ of \eqref{eq:def_jacobian},
                for a scalar system \eqref{eq:renormalized_scalar_equations_set}, initially 
                set for a constant $\ve$, where, eventually, due to limited precision,
                the system moves away from the stationary point towards divergence.
                This is one demonstration where constant $\ve > 0$ cases are \emph{unstable}.
            }
    \label{fig:constant_epsilon_diverging_case_eigenvalues}
\end{figure}

\begin{figure}
    \begin{center}
        \includegraphics[width=14cm]{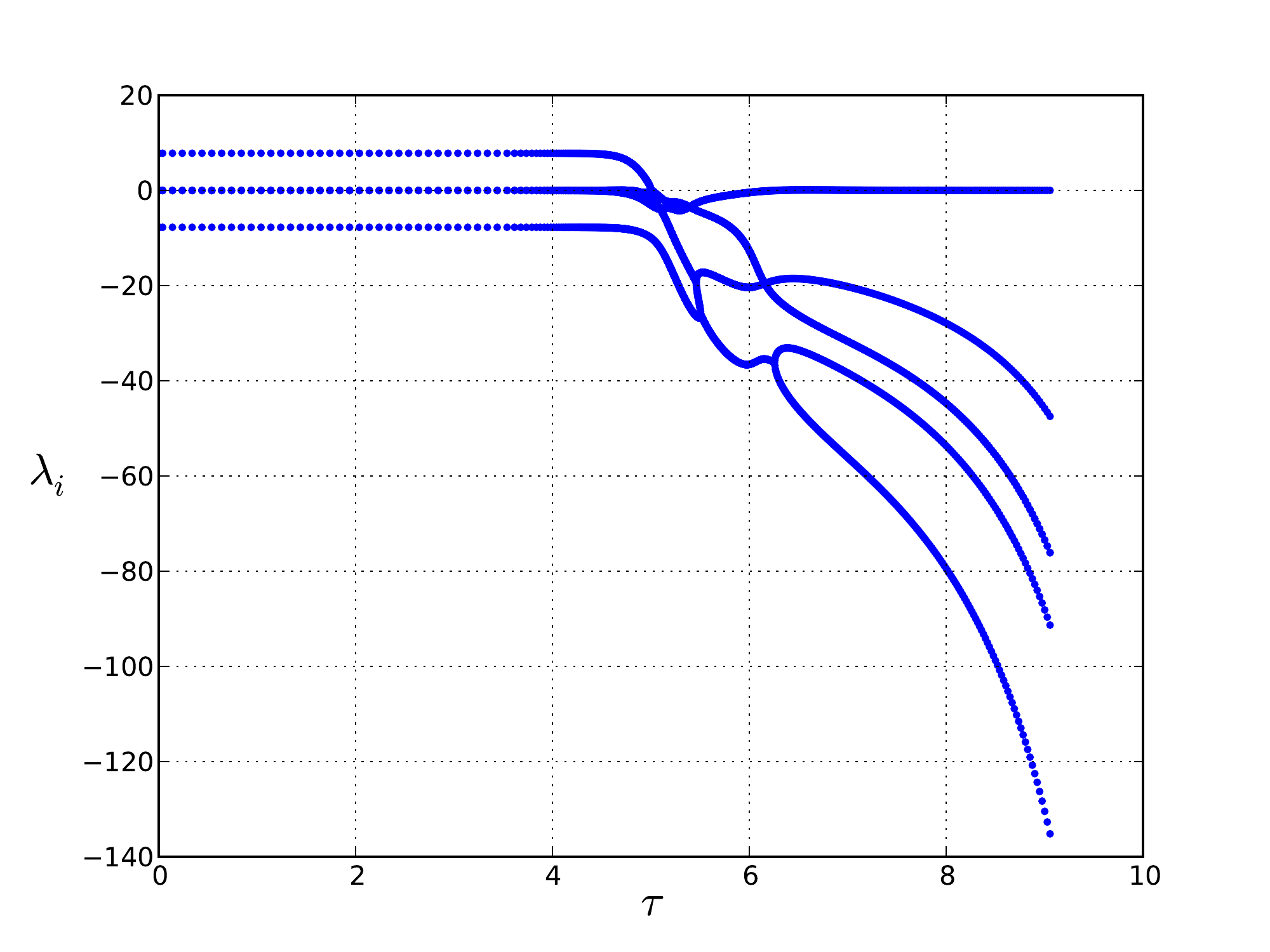}
    \end{center}
    \caption{
                Plot of the $6$ eigenvalues of ${J^{a}}_{b}$ of \eqref{eq:def_jacobian},
                for a scalar system \eqref{eq:renormalized_scalar_equations_set}, initially 
                set for a constant $\ve$, where, in this case, the limited precision leads
                to a convergence towards $\ve \to 0$.
                This is yet another demonstration of the \emph{instability of the constant positive mass-shell deviation}.
            }
    \label{fig:constant_epsilon_converging_case_eigenvalues}
\end{figure}

\begin{figure}
    \begin{center}
        \includegraphics[width=14cm]{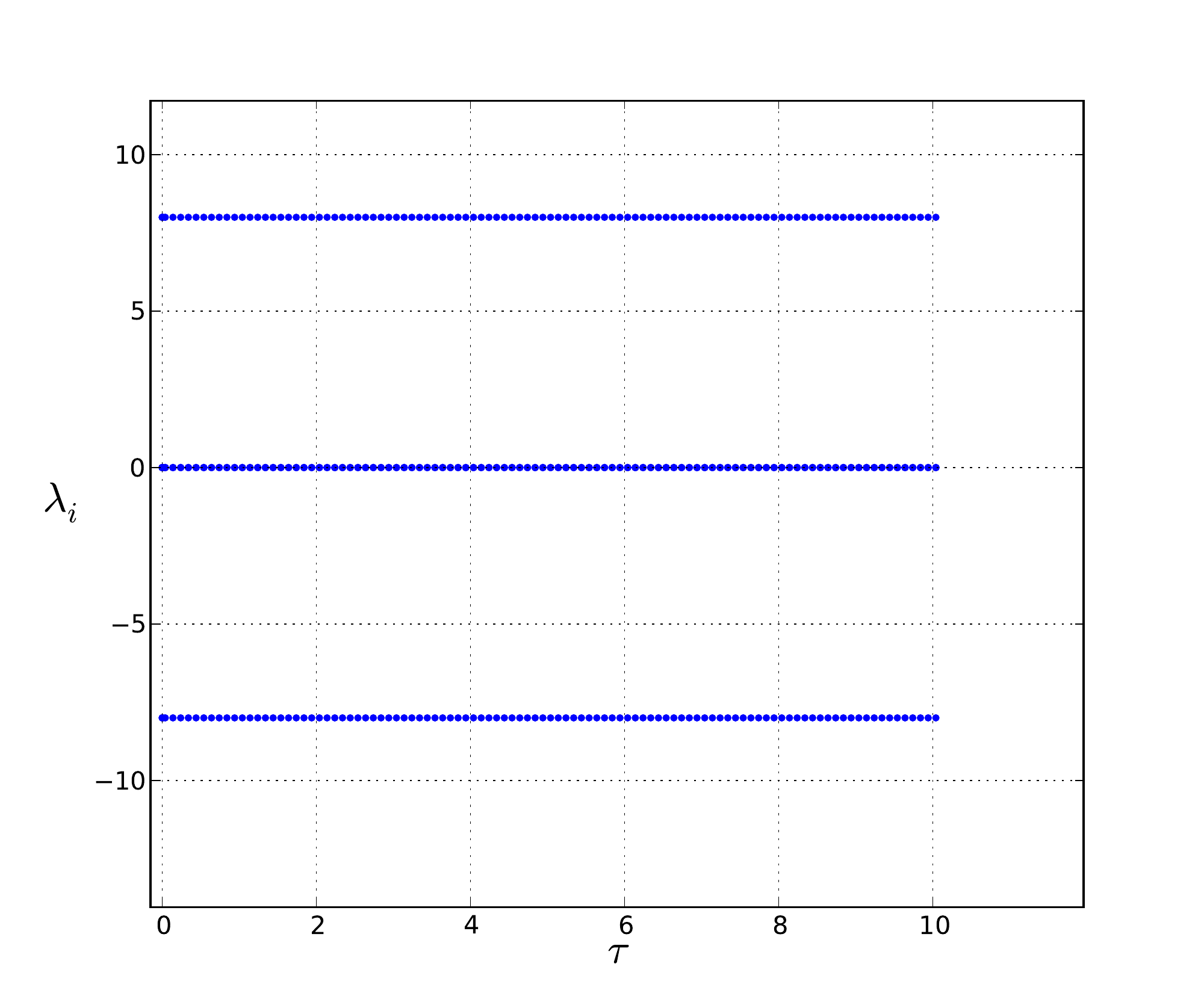}
    \end{center}
    \caption{
                Plot of the $6$ eigenvalues of ${J^{a}}_{b}$ of \eqref{eq:def_jacobian}
                for a scalar system \eqref{eq:renormalized_scalar_equations_set}
                describing a \emph{uniformly moving particle} with a positive
                constant mass-shell deviation.
            }
    \label{fig:constant_epsilon_uniform_motion_eigenvalues}
\end{figure}

To see this, plots of the eigenvalues for an initially \emph{constant $\ve$} are shown in figures 
\ref{fig:constant_epsilon_diverging_case_eigenvalues}, \ref{fig:constant_epsilon_converging_case_eigenvalues}
and \ref{fig:constant_epsilon_uniform_motion_eigenvalues}.

In figures \ref{fig:constant_epsilon_diverging_case_eigenvalues} and \ref{fig:constant_epsilon_converging_case_eigenvalues},
the motion is of an \emph{accelerated particle} such that its $\ve$ is constant 
(see \eqref{eq:renormalized_scalar_equations_set_constant_epsilon}).
Clearly, in both plots, one of the eigenvalues is positive. Limited precision of the numerical simulation
eventually causes the system to shift away from the constant $\ve$, leading either to a 
\emph{diverging $\ve \to \infty$} as in \ref{fig:constant_epsilon_diverging_case_eigenvalues} or 
\emph{converging $\ve \to 0$} as in \ref{fig:constant_epsilon_converging_case_eigenvalues}.
Interestingly, even for \emph{uniform motion} where $\ve > 0$  \ref{fig:constant_epsilon_uniform_motion_eigenvalues},
one of the eigenvalues is positive, which suggests that even this case is unstable under small perturbations.

\subsection{Dependence on $\mmr$}
\label{sec:dependence_on_renormalized_mass}
\begin{figure}
    \begin{center}
        \includegraphics[width=14cm]{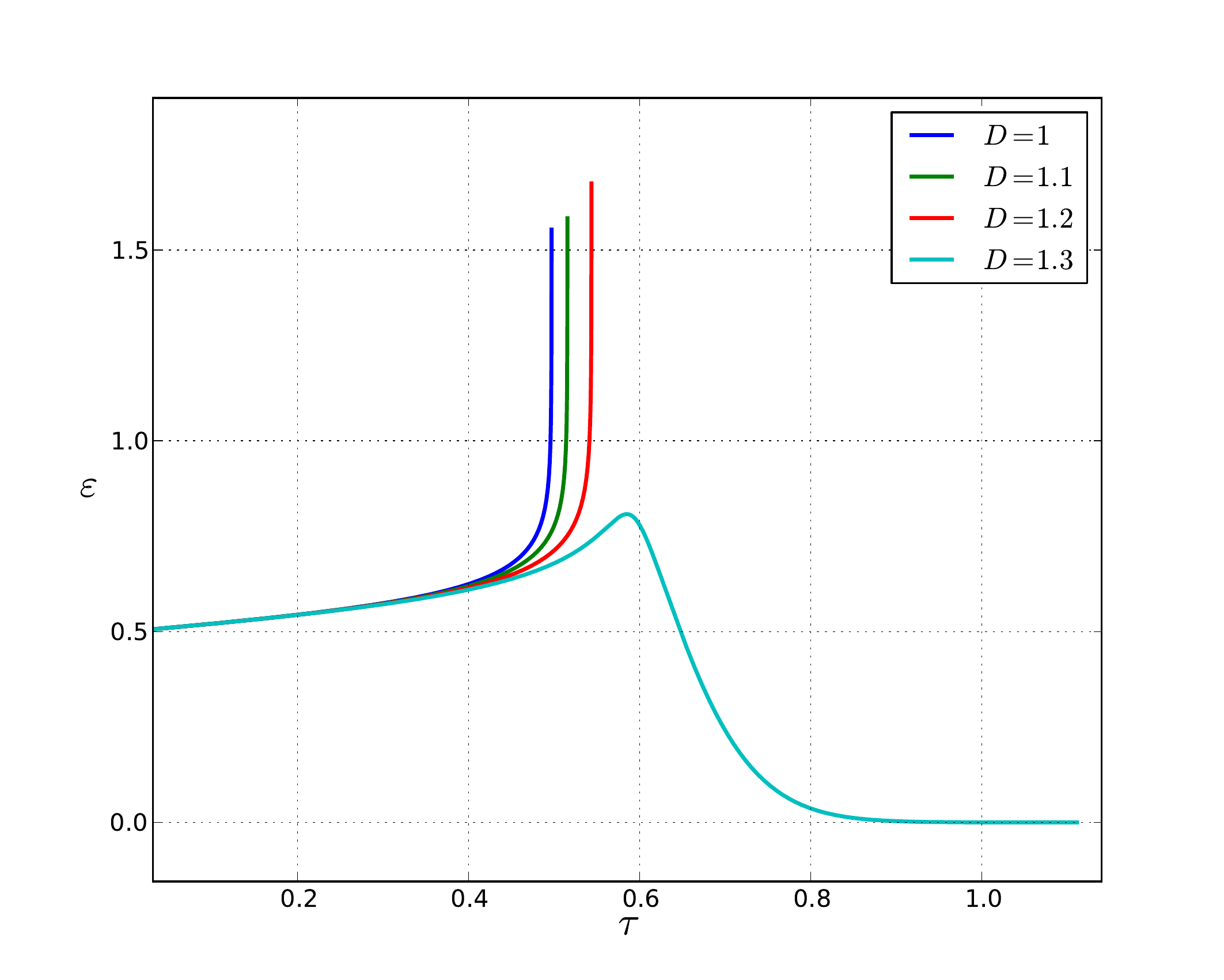}
    \end{center}
    \caption{
                $4$ plots of $\ve(\tau)$, a solution of the scalar system \eqref{eq:renormalized_scalar_equations_set},
                all starting with the \emph{same initial conditions}.
                However, each system has a different parameter $D$ \eqref{eq:renormalized_mass_coefficient}, which 
                scales linearly with the renormalized mass $\mmr$.
                For the lower $3$ values of $D$, the system diverged, and for $D=1.3$, the system converged.
            }
    \label{fig:dependence_on_input_parameter}
\end{figure}

Both equation sets, i.e., equations \eqref{eq:renormalized_lorentz_force_3} and  \eqref{eq:renormalized_scalar_equations_set},
have a single parameter, $D$, defined in \eqref{eq:renormalized_mass_coefficient}, which scales
linearly with the renormalized mass $\mr$.
In figure \ref{fig:dependence_on_input_parameter}, $4$ plots of $\ve(\tau)$ are shown,
each having a different value of $D$.

Thus, it appears as $D \gg 0$, the system becomes more stable, in the sense that conditions which cause 
diverging $\ve \to \infty$ cases, may be tamed to produce converging $\ve \to 0$ by increasing $D$.

At first sight, it may seems contradictory that a large enough $D$ would cause a given set of 
initial conditions to shift from divergence to convergence, as larger $D$ increases $K_2$,
which would attract the system towards divergence even stronger.
However, the key seems to be in the equation for $\dddot{\ve}$ in \eqref{eq:renormalized_scalar_equations_set}.
There, we find, for the $D$ coefficient part
\begin{align}
    \label{eq:dddot_e_focused_on_D_dependence}
    \dddot{\ve}
    & = 
        D \dfrac{\ve^{7/2}}{\ve^2}
        \Big[ 
            - 
            16  \dot{\ve} (\ve + 1) 
            +
            16 \dot{\ve} 
        \Big]
        +
        \ldots
    =
        - 
        16 D \dot{\ve} \ve^{5/2} 
        +
        \ldots
\end{align}
I.e., in this case, larger $D$ would cause a stronger attraction towards \emph{negative} $\dddot{\ve}$.
and therefore, towards lowering $\ve$.

\section{Conclusions and further study}
\label{sec:conclusions}
We have developed the \emph{renormalized Lorentz force} \eqref{eq:renormalized_lorentz_force_3} of a back-reacting particle, 
using the $\tau$-retarded Green-Functions \eqref{eq:tau_retarded_green_function}, first introduced
in \cite{aharonovich_2009}.
The original Lorentz force \eqref{eq:5D_Lorentz_force} was shown to depend on the \emph{sign} of the \emph{mass-shell-deviation}
$\ve$ \eqref{eq:mass_shell_definition_x_dot_x_dot} at and near the present point of the particle $x^{\mu}(\tau)$.
In this paper, we studied the case of $\ve > 0$, in which the particle had a history, possibly short one, 
in which it had a positive mass-shell deviation.
In this case, even though the support of the Green-Function \eqref{eq:tau_retarded_green_function} is the \emph{entire} past 5D lightcone,
the equation was renormalized to avoid infinite force, after which a \emph{local} equation remained.

Note, however, that renormalization is not the only method. 
Other authors have chosen a different subtraction scheme than Gel'fand (see \cite{GalTsov2002} and \cite{Kazinski2002}),
in which the integration on the past remains. 
The method chosen here, however, directly relies on Gal'fand \cite{Gelfand1964_1} regularization method,
and seems more natural,
with the price of mass-renormalization \eqref{eq:renormalized_mass}.
In this sense, this does not differ from other schemes, as even the well studied classical radiation-reaction
in 4D requires mass renormalization if one includes only the retarded part of the 4D Maxwell green-Function
(cf. \cite{Rohrlich1990, Poisson:1999tv, PhysRevD.4.345}).
The subsequent renormalized Lorentz force is $4^{\text{th}}$-order vector equation  \eqref{eq:renormalized_lorentz_force_3}, 
in which $\ddddot{x}^{\mu}$ depends on lower derivatives, and scalars constructed out of them $\ve, \dot{\ve}$ etc.
In the case of no external (renormalized) fields, as was studied in this paper, the vector equation has 12D phase space
constructed of $\dot{x}^{\mu}, \ddot{x}^{\mu}$ and $\dddot{x}^{\mu}$.

Contracting with \eqref{eq:renormalized_lorentz_force_3} with $\dot{x}^{\mu}, \ddot{x}^{\mu}$ and $\dddot{x}^{\mu}$
resulted in a reduced 6D system. This can be understood as follows: the original phase space essentially comprises 
three 4-vectors. We can construct six scalars out of these vectors: the size of each vector, e.g. $\dot{x}_{\mu} \dot{x}^{\mu}$,
and the $3$ \emph{hyperbolic angles} between these vectors, e.g. $\dot{x}_{\mu} \ddot{x}^{\mu}$.
This is, essentially, the reduced scalar system \eqref{eq:renormalized_scalar_equations_set}.

The first scalar, however, was $\ve = - \dot{x}_{\mu} \dot{x}^{\mu} -1$, in which this particular form 
identifies it as a (renormalized) mass-shell deviation $(m^2 - M^2)/M^2$.

Both the vector \eqref{eq:renormalized_lorentz_force_3} and the scalar \eqref{eq:renormalized_scalar_equations_set} equations
where numerically integrated, showing that the system either converges towards the mass-shell $\ve \to 0$, 
or, with more slightly extreme initial conditions, $\ve \to \infty$.
In \emph{both cases}, the 3-velocity $\vb{v} = \vb{\dot{x}} / \dot{t}$ seems to reach a \emph{finite} value,
which indicates that to an observer measuring $\vb{x}$ and $\vb{t}$ of such a particle would see, eventually, moving in uniform motion.
The constant $\ve$ case seems to be unstable, in the sense that small perturbations would 
cause $\ve$ to change, either towards the converging or diverging cases.

The converging $\ve \to 0$, is a classical description of a particle smoothly returning to its own mass-shell
due to self-interaction.
This remarkable result provides a mechanism for the mass stability of a charged particle
in the Stueckelberg framework.
% % % The diverging case, however, implies that the (renormalized) mass of the particle diverges.
% % % Since this can be ruled out as unphysical, a mechanism in which the (renormalized) \emph{Galilean-target-mass}
% % % \emph{increases} is suggested, which is not described in this paper.
As shown in the previous section \ref{sec:dependence_on_renormalized_mass},
if the parameter $D$, which scales linearly with the renormalized mass, 
increases sufficiently, even a diverging system can become convergent.

We have seen that a divergence in $\ve(\tau)$ is accompanied by the existence of a large positive eigenvalue 
(of the Jacobian matrix \eqref{eq:def_jacobian}) for local Lyapunov stability.
A large positive eigenvalue can also have the effect of causing a relatively rapid transition
from a seemingly diverging behavior to an evolution which eventually converges towards the mass shell $\ve \to 0$.

As mentioned above, this paper focused on the \emph{above mass-shell} $\ve > 0$ case, which seems to be consistent, 
in the sense that the self-interaction system maintains $\ve > 0$ even in the converging cases.

The limit point $\ve = 0$, however, is not handled by the above-mass-shell case.
Therefore, in order to complete the study radiation-reaction in this framework, 
one would require a similar development of the Lorentz force \eqref{eq:radiation_reaction_1} 
for the \emph{on-mass-shell} $\ve = 0$ and the \emph{below-mass-shell} $\ve < 0$ cases,
and using \emph{external fields}, describe how a particle could move in and out of these domains.
This will be the subject of a future publication.

% % % % Changing the sign of $\ve$ in effect causes the particle to move in and out of its local 5D light-cone,
% % % % and since the support of the Green-Function is inside the 5D past lightcone,
% % % % the behavior in all these domains is expected to be different.

\appendix

\section{Canonical regularization of divergent integrals}
    \label{sec:canonical_regularization}
%%    \input{UAP_appendix_canonical_regularization}
%%
%%
%%
%%
%%
%%
%%
%%
%% ========================= Appendix A: Canonical regularization ===============================
%%
%%
%%
%%
%%
%%
In this section we provide a short overview of the regularization method described in 
Gel'fand \cite{Gelfand1964_1}.

The function $x^{\lambda}_{+} \equiv \theta(x) x^{\lambda}$ is non-zero for positive $x$, where
$\theta(x)$ is the step-function.
When acting on a smooth bounded function, $\phi(x)$, 
\begin{align*}
    \left( x_{+}^{\lambda} , \phi(x) \right) 
    & = 
        \int_{0}^{\infty}  x^{\lambda} \phi(x) \, \ud x,
\end{align*}
which is well defined for $\Re (\lambda)> -1$. 
On the other hand, the expression can be rewritten as
\begin{align}
    \label{eq:x_lambda_plus}
    \left( x_{+}^{\lambda} , \phi(x) \right) 
    & = 
        \int_{0}^{b}
            x^{\lambda}
            \left[
                \phi(x) 
                -
                \sum_{j=0}^{m}
                    \dfrac{\phi^{(j)}(0)}
                          {j!           }
                    x^{j}
            \right]
            \ud x
        +
        \sum_{j=0}^{m}
            \dfrac{\phi^{(j)}(0)}{j! (\lambda + j + 1)} b^{\lambda + j + 1}
        +
        \nonumber
    \\
    & \qquad
        +
        \int_{b}^{\infty}  x^{\lambda} \phi(x) \, \ud x
\end{align}
where the right-hand-side is well defined for $\{ \Re (\lambda) > -m \} \cap \{ \lambda \neq -1, -2,  \}$.

This suggests that, as a \emph{generalized function}, 
$x_{+}^{\lambda}$ can be \emph{defined} by its action on any smooth bounded function $\phi(x)$,
as given by \eqref{eq:x_lambda_plus}. The result is a function of $\lambda$ defined for 
all $\Re(\lambda) > -m$ except at $\lambda = -1, -2, \ldots -m+1$ where it has \emph{simple poles}
with residues $\dfrac{\phi^{(j)}(0)}{j!}$. This suggests that $x_{+}^{\lambda}$ itself 
is a generalized function with simple poles given by
\begin{align*}
    \text{Res } x_{+}^{\lambda} \Big|_{\lambda = -n}
    & = 
        (-1)^{n}\dfrac{\delta^{(n)}(x)}{n!}
\end{align*}

Similarly, given 2 smooth functions, $\phi(x)$ and $R(x)$, we are seeking 
a regularized solution for
\begin{align}
    \label{eq:R_lambda_definition}
    \left( R_{+}^{-\lambda}(x) , \phi(x) \right)
    & = 
        \int_{a}^{b} \dfrac{\phi(x)}{R^{\lambda}(x)}  \ud x
\end{align}
where, $a$ is defined by $R(a) = 0$, and $R(x) > 0$ for $x \in (a,b)$\footnote{In the meantime, we assume $R(b) > 0$.}.
Furthermore, the expansion of $R(x)$ around $x=a$ begins at some order $m > 0$
\begin{align}
    \label{eq:R_expansion_m_order}
    R(x)
    & = 
        \sum_{n=0}^{\infty}
            \dfrac{(x-a)^{n+m} }
                  {(n+m)!      } 
            A_{n}(a)
        ,
\end{align}
where $A_{n}$ are possible functions of $a$, the lower bound (much like $\tau$ in \eqref{eq:radiation_reaction_1}),
and where we take $A_{0} > 0$, in order for $R(x) > 0$ near $x = a$.

Let us then then define the \emph{remainder function} $T(x)$:
\begin{align}
    \label{eq:T_expansion_definition}
    \begin{split}
        T(x)
        & \equiv
            \dfrac{1}{(x-a)^{m}} R(x)
        =
            \sum_{n=0}^{\infty}
                \dfrac{(x-a)^{n} }
                      {(n+m)!    } 
                A_{n}(a)
    \\
        & = 
            \dfrac{1}{m!} A_{0}(a) 
            + 
            \dfrac{(x-a)}{(m+1)!} A_{1}(a)
            + 
            \ldots
    \end{split}
\end{align}
where $T(a) > 0$, and the upper bound of \eqref{eq:R_lambda_definition} is taken such that $T(x) \neq 0$ throughout $a \leq x \leq b$.
Similarly, the expansion for $\phi(x)$ around $x=a$ is assumed to start at some order $l \geq 0$
\begin{align}
    \label{eq:phi_expansion}
    \phi(x)
    & =
        \sum_{n=0}^{\infty}
            \dfrac{(x-a)^{n+l} }
                  {(n+l)!      } 
            B_{n}(a)
\end{align}
and similarly, the remainder function
\begin{align}
    \label{eq:psi_expansion_definition}
    \psi(x)
    & \equiv
        \dfrac{\phi(x)}{(x-a)^{l}} 
    =
        \dfrac{1}{l!} B_{0}(a) 
        + 
        \dfrac{(x-a)}{(l+1)!} B_{1}(a)
        + 
        \ldots
\end{align}

The integral \eqref{eq:R_lambda_definition} then becomes
\begin{align}
    \label{eq:R_lambda_definition_after_T_definition_1}
    \left( R_{+}^{-\lambda}(x) , \phi(x) \right)
    & = 
        \int_{a}^{b}
            \dfrac{(x-a)^{l        } \cdot \psi(x)        }
                  {(x-a)^{m \lambda} \cdot T^{\lambda}(x) }
            \ud x
    = 
        \int_{a}^{b}
            \dfrac{\psi(x)              }
                  {T^{\lambda}(x)       }
            \dfrac{\ud x}{(x-a)^{m \lambda - l}}
\end{align}
Defining
\begin{align}
    \label{eq:q_def}
    q(x)
    & \equiv
        \dfrac{\psi(x)          }
              {T^{\lambda}(x)   }
\end{align}
we then have
\begin{align}
    \label{eq:R_lambda_definition_after_T_definition_2}
    \left( R_{+}^{-\lambda}(x) , \phi(x) \right)
    & = 
        \int_{a}^{b}
            \dfrac{q(x)}
                  {(x-a)^{m \lambda - l}}
            \ud x
\end{align}

In case $m\lambda-l \in \{ 1/2, 3/2, \ldots\}$ is non-integer, 
then the regularization given in \eqref{eq:x_lambda_plus} 
is well defined, where we take $x_{+}^{\lambda} \to (x-a)_{+}^{m \lambda - l}$ 
and $\phi(x) \to q(x)$.

When $m\lambda - l \in \mathbb{N}$ is an integer, 
then regularization no longer applies, and the result has a simple pole 
whose residue is given by
\begin{align}
    \label{eq:residue_R_lambda_definition_after_T_definition_2}
    \text{Res}
    \left( R_{+}^{-\lambda}(x) , \phi(x) \right)
    \Big|_{m \lambda - l \in \mathbb{N}}
    & = 
        b^{\lambda}
        (-1)^{m\lambda - l - 1}
        \dfrac{\ud^{m\lambda - l - 1} }
              {\ud x^{m\lambda - l - 1}}
        q(x)
        \Big|_{x = a}
\end{align}

For the above-mass-shell deviation, e.g., as given in \eqref{eq:radiation_reaction_4},
we find $\lambda = 5/2, m = 2$ and $l = 2$, and therefore, $m\lambda - l = 3$,
which residue is the renormalized force given in \eqref{eq:radiation_reaction_renormalized_1}.

\newpage
\bibliographystyle{amsplain}
\bibliography{bibliography-all}

\end{document}